\documentclass[%
 reprint,
 amsmath,amssymb,
 aps,
]{revtex4-2}

\usepackage{hyperref}
\hypersetup{
    colorlinks=true, 
    linkcolor=cyan,
    citecolor=magenta, 
    filecolor=magenta, 
    urlcolor=cyan,
    runcolor=cyan
}

\UseRawInputEncoding
\usepackage{hyperref}
\usepackage{graphicx}
\usepackage{dcolumn}
\usepackage{bm}
\usepackage{subfig}
\usepackage{color,soul}
\usepackage{float}
\usepackage{mathtools}
\usepackage{siunitx}
\usepackage{booktabs}

\usepackage[english]{babel}

\makeatletter
\renewcommand{\fnum@figure}{FIG. \thefigure}
\makeatother

\begin{document}
\title{Modeling betatron radiation using particle-in-cell codes for plasma wakefield accelerator diagnostics}

\author{M. Yadav$^{1,2,3}$}
\email{monika.yadav@liverpool.ac.uk}
\author{C. Hansel$^{4}$}
\author{B. Naranjo$^{1}$}
\author{G. Andonian$^{1}$}
\author{P. Manwani$^{1}$}
\author{{\"O.} Apsimon$^{5}$}
\author{C.P. Welsch$^{2,3}$}
\author{J. B. Rosenzweig$^{1}$}

\affiliation{$^1$Department of Physics and Astronomy, University of California Los Angeles, California 90095, USA}

\affiliation{ 
$^2$ Department of Physics, University of Liverpool, Liverpool L69 3BX, UK
 }
 \affiliation{ 
   $^3$ Cockcroft Institute, Warrington WA4 4AD, UK
 }

\affiliation{$^4$Center for Integrated Plasma Studies, Department of Physics, University of Colorado Boulder, Boulder, Colorado 80309, USA}

\affiliation{ 
 $^5$Department of Physics and Astronomy, University of Manchester, Manchester M19 3PL, UK
 }

\date{\today}

\begin{abstract}


The characterization of plasma wakefield acceleration experiments using emitted photons from betatron radiation requires numerical models in support of instrumentation of single-shot, double-differential angular-energy spectra. Precision characterization for relevant experiments necessitates covering a wide energy range extending from tens of keV through 10~GeV, with an angular resolution on the order of $100 \;\mu\text{rad}$.  
In this paper, we present a numerical model for betatron radiation from plasma accelerated beams, that are based on integration of Liénard–Wiechert (LW) potentials for computed particle trajectories. The particle trajectories are generated in three different ways: first, by particle tracking through idealized fields in the blowout regime of PWFA; second, by obtaining trajectories from the Quasi-static Particle-in-Cell (PIC) code \texttt{QuickPIC}; and third, by obtaining trajectories from the full PIC code \texttt{OSIRIS}. We performed various benchmarks with analytical expressions and the PIC code \texttt{EPOCH}, which uses a Monte-Carlo QED radiation model. Finally, we present simulations of the expected betatron radiation for parameters from the Facility for Advanced Accelerator Experimental Tests II (FACET-II) PWFA and plasma photocathode experiments.
\end{abstract}

\maketitle

\section{Introduction}

Beam driven plasma wakefield accelerators (PWFAs) have already demonstrated \cite{blumenfeld2007} acceleration gradients orders of magnitude higher than the limits of conventional radio-frequency acceleration and are envisioned as the core technology for next-generation, compact high-energy particle accelerators. In PWFA, a drive electron beam in a plasma excites a wakefield that accelerates a trailing (witness) electron beam.
The Facility for Advanced
Accelerator Experimental Tests (FACET) at SLAC National Accelerator Laboratory \cite{Hogan2010}, the predecessor facility to FACET-II, achieved several key milestones toward realizing practical PWFAs including high-efficiency acceleration \cite{Litos2014}, high total energy gain \cite{Litos_2016}, positron acceleration \cite{Corde2015}, and the demonstration of a plasma photocathode \cite{Deng2019}, also known as Trojan horse. 
The FACET-II facility \cite{facet2,Joshi_2018} aims to build on these accomplishments by demonstrating high beam quality of accelerated and plasma-injected beams and testing several new concepts for acceleration, manipulation, and radiation generation. 
However, measuring many key results of these experiments is a significant challenge for conventional diagnostics, due to the extremely low emittances, potentially large correlated energy spread, and jitter of the anticipated beams. 
While using advanced machine learning (ML) methods \cite{ml1,ml2,ml3} are viable for predicting the longitudinal phase-space, non-destructive inference of transverse beam emittance, and spectral reconstruction of the bunch profile, electron beam facilities will also rely on betatron radiation as a critical mechanism to augment the diagnostic capacity.

In a PWFA, betatron radiation \cite{esarey,Sebastian_2013,kostyukov2003} is produced due to the transverse betatron oscillations of charged particles. Betatron radiation is beneficial not only as a radiation source, but also as a carrier of detailed information about beam-plasma interaction. Employing betatron radiation as a diagnostic has the advantage of being both single-shot and non-destructive to the beam. 
The betatron radiation signal is accessible due to the large radiation flux generated by relativistic beams in plasmas. Betatron radiation diagnostics have been utilized in inverse Compton scattering experiments \cite{sakai2017single}, and studied in proton-driven PWFA experiments at AWAKE \cite{Williamson, awake_2022}. While the implementation of betatron radiation diagnostics was attempted at FACET, the high emittance of the beams prevented it from meeting its full potential. The betatron diagnostic system will provide both angular and spectral information about the emitted betatron radiation on a shot-by-shot basis, yielding critical information about the plasma accelerated beam.



In order to practically infer information about the beam from betatron radiation, fast and accurate numerical models are required. Betatron radiation models should generate large datasets for ML, or used within a maximum likelihood estimation algorithm (MLE), or similar. Recent work \cite{ML-MLE-2022,zhuang1,yadav_2022,yadav2021_AAC,Maanas} has utilized data from simulated PWFA betatron radiation experiments to reconstruct beam parameters using ML methods reliably. However, the radiation spectrum in this preliminary analysis extends beyond 100 MeV due to the very high plasma density and concomitant focusing strength. Studies of the changes to this spectrum due to instabilities and more significant amplitude betatron motion are now underway.


This work discusses novel radiation algorithms to study the radiations emitted by beam plasma interactions. There are various techniques to compute the motion of charged particles, including analytical methods mathematically modeling the charged particle's motion using equations of motion and solving them analytically. Numerical integration using Runge-Kutta method, computational approach to model charged particle dynamics in plasma by treating particles as discrete entities and solving equations of motion in a self-consistent manner. A statistical Monte Carlo simulation technique for simulating the motion of charged particles by randomly sampling their path through space and averaging the results over many runs. 
In this paper, we introduce numerical models for computing betatron radiation spectrum from PWFAs and plasma injectors that can be used to reconstruct beam parameters from betatron radiation signatures. Radiation is computed based on integration of LW potentials for computed particle trajectories using different PIC codes. The radiation spectra for different models are validated, compared, and used to simulate expected radiation properties from planned PWFA and plasma injector experiments at FACET-II. Spatial and temporal profiles of radiation are also important to study orbital angular momentum of light, microscopy of lights \cite{twistedlight}.

In Sec. \ref{sec:analytic}, we lay out a limited analytical description of the betatron radiation generated by a Gaussian beam in an ion channel. In Sec. \ref{sec:models}, we discuss numerical models for betatron radiation from plasma accelerated beams for computed particle trajectories for different PIC code. In Secs. \ref{sec:pwfa} and \ref{sec:photocathode}, we present simulations of planned PWFA and plasma injection experiments, respectively, and discuss generally expected features of the radiation. Finally, in Sec. \ref{sec:conclusion}, we conclude with numerical modeling of betatron radiation and betatron radiation diagnostics in general.

\section{\label{sec:analytic} Analytic description of the radiation spectrum}

In this section, we present an analytic model of betatron radiation in PWFA, beginning with the single-particle spectrum derivation and extending it to a particle distribution. In the blowout regime \cite{jamie_blowout}, the drive beam leaves behind an ion channel that can be assumed uniform provided ion motion \cite{jamieionmotioncalc} is insignificant. For paraxial ($\bm{p}_{\perp} \ll p_z$) motion, beam electrons undergo simple harmonic motion with a betatron angular wavenumber, $k_{\beta} = k_p / \sqrt{2 \gamma}$, where $\gamma$ is the Lorentz factor, $k_p = \sqrt{4 \pi r_e n_0}$ is the plasma angular wavenumber, $n_0$ is the plasma density, and $r_e$ is the classical electron radius \cite{kostyukov2003}. Beam electrons produce undulator radiation with an equivalent strength parameter $K = \gamma k_{\beta} r_{\beta}$, where $r_{\beta}$ is the particle's maximum oscillation amplitude.


In the ion channel, each particle of an on-axis beam has a different value of $K$, due to different oscillation amplitudes. Contrasted with undulator radiation, where $K$ is constant, the spectra of PWFA betatron radiation crosses different regimes. There are three regimes of undulator radiation \cite{Sebastian_2013}. For $K \ll 1$, radiation is emitted in a cone containing angles $\theta \lesssim 1 / \gamma$, where $\theta$ is the radiation cone opening angle. Then, the radiation spectrum is sharply spiked around the total photon energy $\epsilon_1 = 2 \hbar c \gamma^2 k_{\beta} / (1 + \gamma^2 \theta^2 + K^2 / 2)$. Then, for $K \sim 1$, integer harmonics of the fundamental begin to be generated, and the radiation is emitted in the broader cone containing angles $\theta \lesssim K / \gamma$. More harmonics are produced for larger $K$. Finally, for $K \gg 1$, the harmonics blend to form a smooth synchrotron radiation spectrum characterized by a critical photon energy $\epsilon_c = 3 \hbar c K \gamma^2 k_{\beta} / 2$.

These three regimes describe the spectra produced by a single particle. The total radiation is given by the integral over a range of $K$ for a beam. Provided the beam spot size $\sigma_r \gg 1 / (\gamma k_{\beta})$, many of the particles in the beam are in the $K \gg 1$ regime, and since those particles generate significantly greater number of photons than a particle with small $K$, the overall radiation emitted by the beam is dominated by them.

The radiation of a single particle in the $K \gg 1$ regime is, assuming no $\bm{\hat{z}}$ angular momentum,

\begin{equation}
\left(\frac{dI}{d\epsilon}\right)_{sp} = \frac{I_{\mathrm{tot}, sp}}{\epsilon_{c,sp}} S_{sp} \left( \frac{\epsilon}{\epsilon_{c,sp}} \right)
\end{equation}

\noindent where $\epsilon_{c,sp}$ and $I_{\mathrm{tot}, sp}$ are the critical photon energy and total radiated energy by a single particle, respectively which is given by \cite{esarey}

\begin{equation}
I_{\mathrm{tot}, sp} = \frac{1}{6} r_e m_e c^2 k_p^4 L_p \gamma^2 r_{\beta}^2,
\end{equation}

\begin{equation} \label{eq:ecsingleparticle}
\epsilon_{c, sp} = \frac{3}{4} \hbar c k_p^2 \gamma^2 r_{\beta},
\end{equation}

\noindent and 

\begin{equation} \label{eq:sxsingleparticle}
S_{sp}(x) = \frac{9 \sqrt{3}}{8 \pi} x \int_x^{\infty} K_{5/3}(y) dy
\end{equation}

\noindent where $K$ is the Bessel function. $S_{sp}(x)$ is the universal function of synchrotron radiation \cite{Sebastian_2013}, which satisfies the normalization condition $\int_0^{\infty} S_{sp}(x) dx = 1$. The single particle spectrum can be integrated over the beam distribution to express the spectrum of the total radiation produced by the beam. For a monochromatic Gaussian beam, the spectrum is

\begin{equation} \label{eq:beamspectrum}
\begin{split}
\left(\frac{dI}{d\epsilon} \right)_{b} &= 2 \pi \int_0^{\infty} \left(\frac{dI}{d\epsilon}\right)_{sp}  \frac{Q}{2 \pi \sigma_{\perp}^2 e} e^{-\frac{r_{\beta}^2}{2 \sigma_{\perp}^2}} r_{\beta} dr_{\beta} \\
&= \frac{I_{\mathrm{tot},b}}{\epsilon_{c,b}} S_b \left(\frac{\epsilon}{\epsilon_{c,b}} \right)
\end{split}
\end{equation}

\noindent where $Q$ is the total beam charge, and $\sigma_\perp$ is the average radius of the beam particles,

\begin{equation} \label{eq:sxbeam}
S_{b}(x) = \frac{9 \sqrt{3} \Lambda^4}{16 \pi} x^3 \int_0^{\infty} \int_{v}^{\infty} \frac{K_{5/3}(u) e^{-\frac{\Lambda^2 x^2}{2 v^2}}}{v^3} du dv,
\end{equation}

\begin{equation}
I_{\mathrm{tot},b} = \frac{1}{3 e} r_e m_e c^2 k_p^4 L_p Q \gamma^2 \sigma_r^2,
\end{equation}

\noindent and

\begin{equation} 
\epsilon_{c,b} = \frac{3}{4} \Lambda \hbar c k_p^2 \gamma^2 \sigma_r
\end{equation}

\begin{figure}[h!]
    \centering
    \includegraphics[width=\columnwidth]{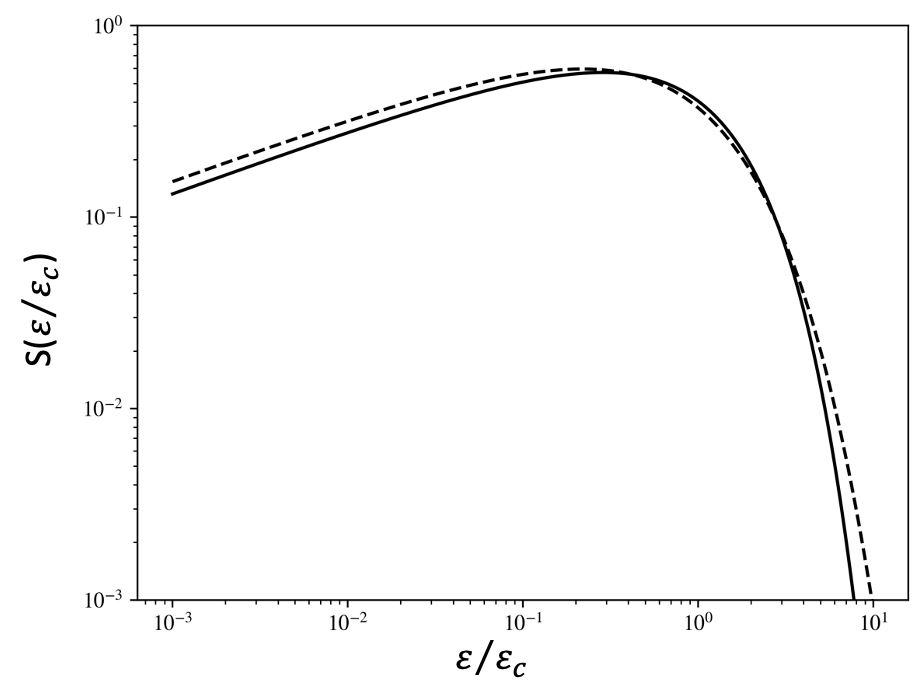}  \includegraphics[width=\columnwidth]{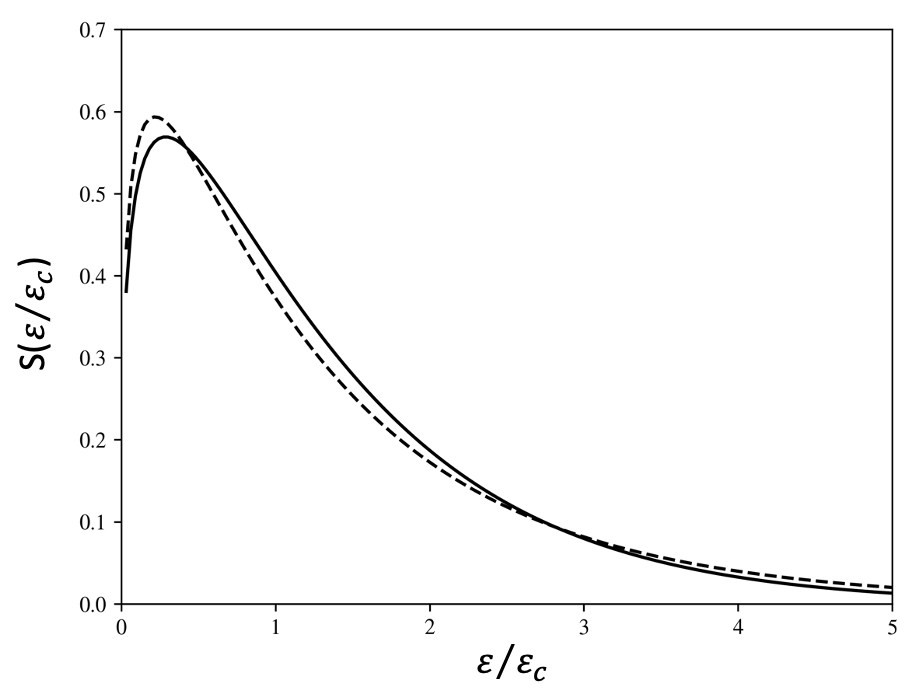}
    \caption{Normalized analytic betatron radiation spectra as functions of the normalized radiation energy. Solid line: Single particle spectrum $S_{sp}(\epsilon / \epsilon_c)$ given by equation (\ref{eq:sxsingleparticle}). Dashed line: Beam spectrum $S_b(\epsilon / \epsilon_{c,b})$ given by Eq. (\ref{eq:sxbeam}).}
    \label{fig:sanalytic}
\end{figure}

\noindent where $\Lambda$ is a dimensionless constant determined using the precise definition of the critical energy. We have defined $S_b$(x) such that it satisfies the same normalization condition as $S_{sp}(x)$: $\int_0^{\infty} S_b(x) dx = 1$. The critical energy $\epsilon_c$ is defined as the energy for which \cite{hofmann}

\begin{equation}
\int_0^{\epsilon_c} \frac{d I}{d \epsilon} d \epsilon = \int_{\epsilon_c}^{\infty} \frac{d I}{d \epsilon} d \epsilon.
\end{equation}

\noindent In order to define $\epsilon_{c,b}$ such that this is satisfied in the beam case, the constant $\Lambda$ must satisfy

\begin{equation}
\int_0^{\infty} \left[1 - e^{-\frac{\Lambda^2}{2 x^2}} \left(1 + \frac{\Lambda^2}{2 x^2}\right) \right] S(x) dx = \frac{1}{2}.
\end{equation}

\noindent Numerically evaluating this gives $\Lambda \approx 1.7231$. A plot of $S_{sp}(x)$ and $S_b(x)$ is shown in Fig. \ref{fig:sanalytic}. From this plot, it is clear that the overall shape of the normalized single particle and beam spectra $S_{sp}(x)$ and $S_b(x)$ are similar. Overall while these equations are of some usefulness, they do not account for many of the important effects that may influence betatron radiation, such as acceleration, energy spread, $\bm{\hat{z}}$ angular momentum of beam electrons, plasma ramps, and the contribution of the low $K$ core of the beam.

\section{Numerical Models of Betatron Radiation}\label{sec:models}

This section presents three numerical models for computing betatron radiation at three increasing fidelity, and computational cost, levels. Higher grid resolution is required to capture short wavelength radiation using PIC algorithms, leading to high computational expensive codes.

\subsection{Idealized particle tracker with Liénard–Wiechert radiation} \label{sec:model1}

We developed a betatron radiation code that tracks particles through idealized fields and computes the emission of electromagnetic radiation by charged particles undergoing betatron oscillations using Liénard–Wiechert (LW) potentials. The code is written in C++ and parallelized using $\texttt{Boost.MPI}$ allowing for large simulations. First, macro-particles are randomly sampled from a Gaussian beam distribution. Next, particles are tracked through idealized acceleration and focusing fields using a 4th-order Runge-Kutta (RK4) integration method. The fields used are a linear focusing force from the ion channel and a constant accelerating field $\bm{E} = Z_i m_e \omega_p^2 \bm{r}_{\perp} / 2 e + E_{\mathrm{accel}} \bm{\hat{z}}$ where $E_{\mathrm{accel}}$ is an input parameter. The electron trajectories are used to integrate the complex LW potential for particle $i$ numerically.

\begin{equation} \label{eq:lw1}
 \bm{V}_i =  \int_{t_i}^{t_f} \frac{\bm{n} \times (( \bm{n} - \bm{\beta} ) \times \dot{\bm{\beta}} )}{(1 - \bm{n} \cdot \bm{\beta})^2} e^{\frac{i \epsilon}{\hbar} ( t - \bm{n} \cdot \bm{r}(t) / c )} dt
\end{equation}

\begin{figure*}
    \centering
    \includegraphics[width=1\textwidth]{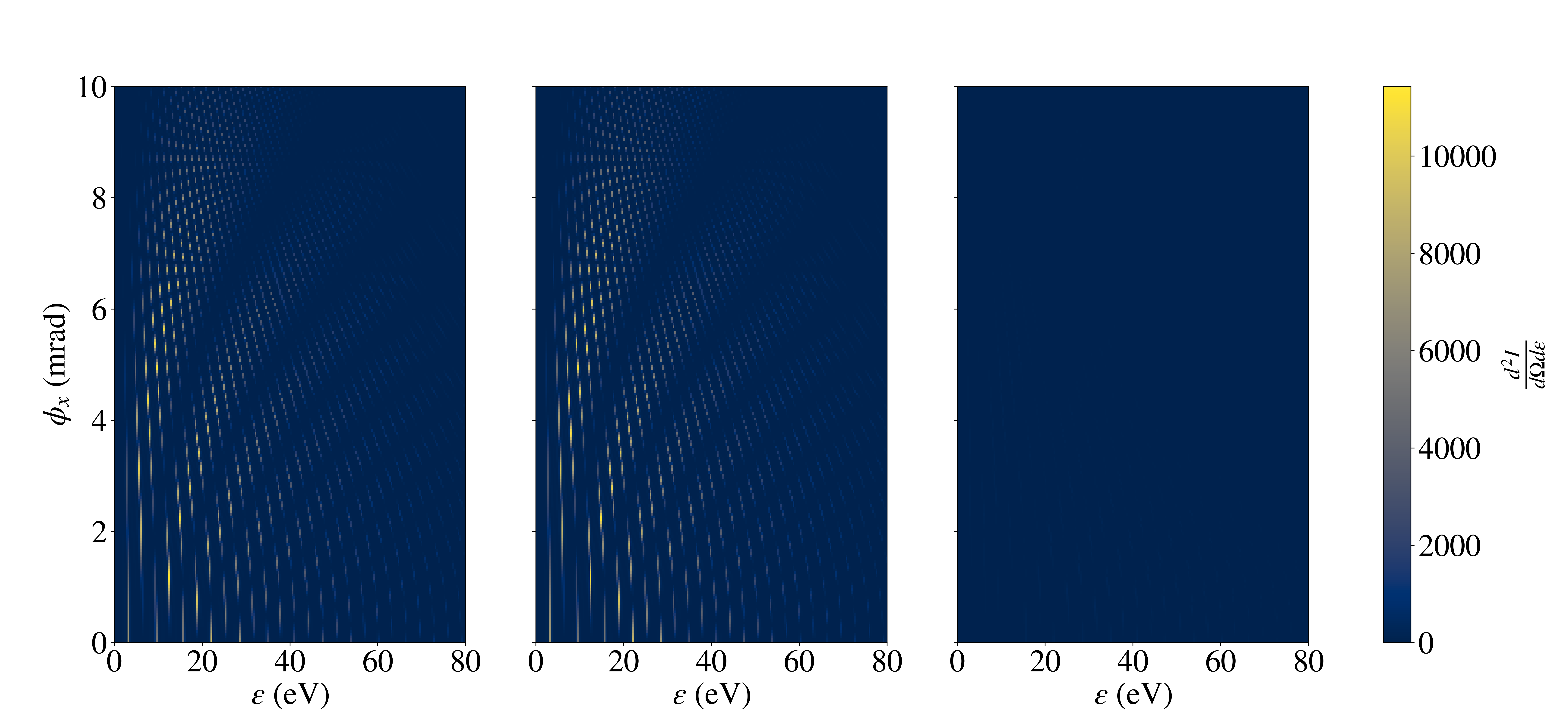}
    \caption{Double differential spectra for the second benchmark of the model \ref{sec:model1}. Left: Analytical spectrum. Middle: Spectrum computed by a model \ref{sec:model1}. Right: Absolute error between the two spectra.}
    \label{fig:model1validation1}
\end{figure*}

\begin{table}
    \centering
    \caption{Parameters for the second benchmark simulation for the idealized particle tracker and LW. \label{tab:model1validation2params}}
    \begin{ruledtabular}
    \begin{tabular}{ccc}
    Parameter & Value & Unit \\
    \hline
    Plasma density & $4\times 10^{16}$ & cm$^{-3}$ \\
    Plasma length & $60$ & cm \\
    Beam energy & $10$ & GeV \\
    Beam charge & $500$ & pC \\
    Beam spot size & $4.5$ & $\mu$m \\
    Beam normalized emittance & $0$ or $75$ \footnote{Two simulations were performed, one with zero emittance and one with matched emittance.} & mm-mrad \\
    Simulation particles & $500$ & \\
    Step size & $12$ & $\mu$m \\
    $\phi_{x,y}$ window & $[-1.5,1.5]$ & mrad \\
    $\phi_{x,y}$ points & 51 & \\
    $\epsilon$ range & $[0.5,5000]$ & keV \\
    $\epsilon$ points & 101 & \\    
    $\epsilon$ spacing & logarithmic & \\
    \end{tabular}
    \end{ruledtabular}
\end{table}

\noindent where $\bm{n}$ is the average vector in the direction of the radiation observation direction, ${\bm{\beta}}$ is the velocity of the electron normalized to the speed of light c, $\dot{\bm{\beta}}$ is the usual acceleration divided by speed of light and $\epsilon$ is the photon energy. $\bm{V}_i$ is computed over a 3D grid of different directions and photon energies. This computation takes place simultaneously with particle tracking, the particle data does not have to be saved and can be discarded after it has been used to compute the fields. Each Message Passing Interface (MPI) process computed its contribution to the radiation independently and in parallel and summed at the end. The double differential spectrum is given by

\begin{equation} 
\frac{d^2 I}{d\Omega d\epsilon} = \frac{e^2}{16 \pi^3 \epsilon_0 \hbar c} \left | \sum_i w_i \bm{V}_i \right |^2
  \label{eq:sumrad}
\end{equation}

\noindent where $w_i \equiv \sqrt{|Q_{i, \mathrm{macroparticle}}| / e}$ is the particle weight. The primary bottleneck is the $O(N^3)$ scaling of the 3D grid resolution and the small time step required to prevent aliasing, particularly at high $\epsilon$. While the number of simulated particles is typically small, even with just a few hundred, the statistical error in computed radiation tends to be low.

 \begin{figure}
    \centering
    \includegraphics[width=\columnwidth]
    {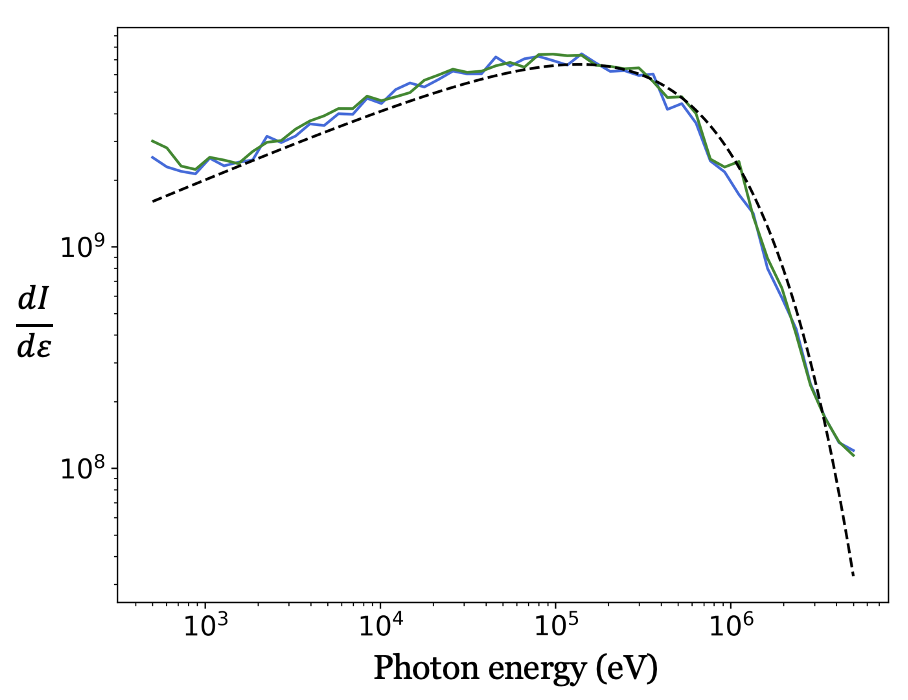}
    \caption{Synchrotron radiation spectra for the second benchmark of the model \ref{sec:model1}. Blue: simulation with zero emittance. Green: simulation with matched emittance. Black, dashed: analytic spectrum.}
    \label{fig:model1validation2}
\end{figure}

We performed two tests to benchmark the code. We tracked a single particle through an ion channel in the first test. The particle had an energy $E = 100\, \mathrm{MeV}$, undulator parameter $K = 2$, and betatron period $\lambda_{\beta} = 1\, \mathrm{cm}$. It was tracked for $10$ periods with $100$ steps per period. The analytical and numerical double differential spectra and the absolute error between them are shown in Fig. \ref{fig:model1validation1}, demonstrating agreement between simulation and theory. For the second benchmark, we tracked a beam with parameters roughly based on PWFA experimental plans at FACET-II, which are shown in Table \ref{tab:model1validation2params}. Two simulations were performed, one with zero emittance and one with finite emittance, such that the spot size was matched to the plasma $\epsilon = \gamma k_{\beta} \sigma_r^2$. The spectra from both of these simulations are shown compared to the analytic expression from Eq.~(\ref{eq:beamspectrum}) in Fig. \ref{fig:model1validation2}.

\subsection{Quasi-Static Particle-in-Cell with Liénard Wiechert Radiation} \label{sec:model2}

At the next level of sophistication, we computed betatron radiation from the 3D Quasi-static PIC code \texttt{QuickPIC} \cite{qpic1,qpic2}. Quasi-static PIC codes use the approximation that the beam evolves on a much slower timescale than the plasma wake to achieve significant speedups over conventional PIC codes. This approximation is accurate when simulating PWFA but prevents Quasi-static from simulating plasma injectors or non-relativistic beams. Compared to that of Sec. \ref{sec:model1}, this numerical model can accurately describe the radiation from the drive beam, which is only partially inside the ion channel, and it can accurately describe the radiation signature of effects such as hosing. While $\texttt{QuickPIC}$ does not directly compute radiation, we modified it to output particle trajectories and input a randomly selected subset of those into a code based on the LW code discussed in Sec. \ref{sec:model1}. This approach is similar to the method used in \cite{PSan} to compute betatron radiation.

\begin{table}
    \centering
    \caption{Parameters for \ref{sec:model2} benchmark simulation}
    \label{tab:model2validation1params}
    \begin{ruledtabular}
    \begin{tabular}{ccc}
    Parameter & Value (Drive, Witness) & Unit \\
    \hline
    Plasma density & $4\times 10^{16}$ & cm$^{-3}$ \\
    Plasma length & $60$ & cm \\
    Plasma radius & $31.9$ & $\mu$m \\
    Beam energy & $10$, $10$ & GeV \\
    Beam charge & $500$, $500$ & pC \\
    Beam spot size & $5$, $4.5$ & $\mu$m \\
    Beam length & $5$, $2.8$ & $\mu$m \\
    Beam norm. emit. & $3.2$, $3$ & mm-mrad \\
    Beam separation & $101.55$ & $\mu$m \\
    \texttt{QuickPIC} Resolution & $1.04 \times 1.04 \times 0.60$ & $\mu$m \\
    \texttt{QuickPIC} Time step & $1.10$ & ps \\
    \texttt{QuickPIC} Macro-particles & $64^3$ & \\
    LW particles & $100$ & \\
    LW time step & $66.7$ & fs \\
    LW angular window & $[-1, 1]$ & mrad \\
    LW angular grid points & $25 \times 25$ & \\
    LW angular grid spacing & linear & \\
    LW energy window & $[5 \times 10^3, 5 \times 10^6]$  & eV \\
    LW energy grid points & $50$ & \\
    LW energy grid spacing & logarithmic & \\
    \end{tabular}
    \end{ruledtabular}
\end{table}

\begin{figure}[h!]
    \centering
    \includegraphics[width=\columnwidth]{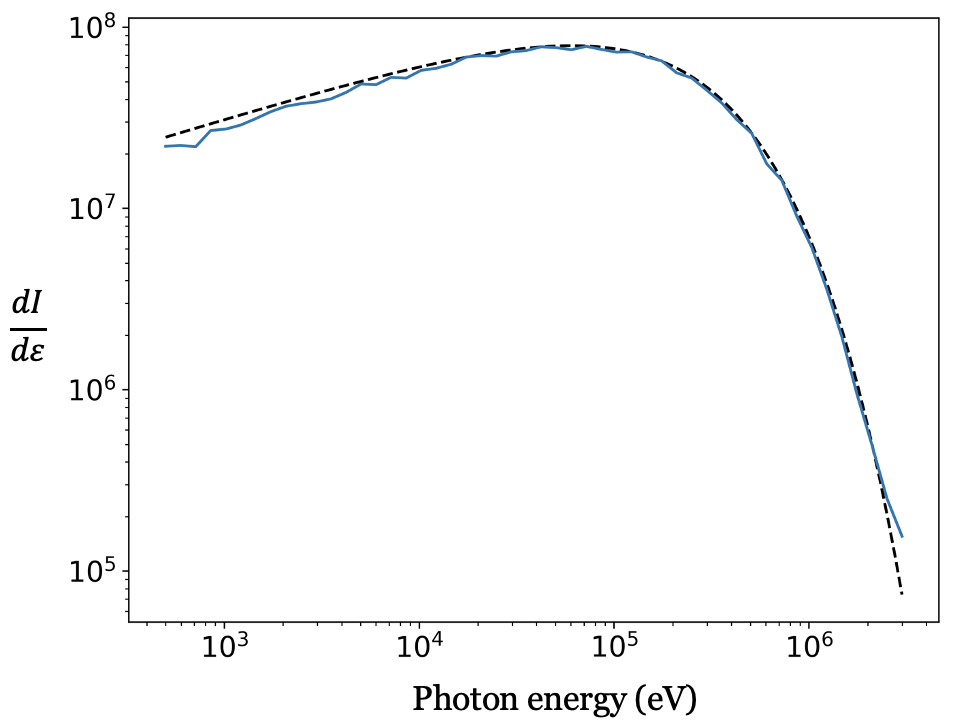}
    \caption{Blue: Radiation spectrum computed numerically using the \ref{sec:model2}. Black, dashed: Analytical radiation spectrum given by Eq. (\ref{eq:beamspectrum}).}
    \label{fig:model2validation1}
\end{figure}

When computing the LW potential integral Eq.~(\ref{eq:lw1}), the step size required to accurately compute high energy radiation without aliasing effects is very small, typically much smaller than the step size required to track particles accurately. In order to compute this high energy radiation, additional trajectory points were interpolated between the points computed by \texttt{QuickPIC} using cubic B-spline interpolation. This was not done in Sec. \ref{sec:model1} because the speed of the simple RK4 tracker in Sec. \ref{sec:model1} meant that particles could be tracked with smaller steps than needed while barely increasing computation time. Additionally, the LW code used in this section used Python's $\texttt{multiprocessing}$ module, and computed radiation after particle tracking finished.

In order to benchmark this code against Eq.(\ref{eq:beamspectrum}), we used parameter set, shown in Table \ref{tab:model2validation1params}, that minimized witness beam acceleration by placing it at the longitudinal wakefield's zero crossing. Only the radiation from the witness beam was computed, and the time evolution of the drive beam was turned off. The computed spectrum, shown in Fig. \ref{fig:model2validation1}, agrees with the theoretical spectrum.

\subsection{Full Particle-in-Cell OSIRIS code with Liénard–Wiechert Radiation} \label{sec:model3}



We will use the \texttt{OSIRIS} code to simulate the FACET-II Trojan horse experimental scenario. These simulations focuses on generating the witness beam using plasma photocathodes and collinear laser ionization injection scheme. In the end, we will focus on calculating betatron radiation importing trajectories obtained from \texttt{OSIRIS} to Sec. \ref{sec:model1}. A plasma profile comprising a vacuum section followed by an increasing short ramp section was implemented. The zero density section is provided for the initialization of a laser pulse by ensuring the consistency of Maxwell's equations. A laser pulse is used in \texttt{OSIRIS} code defining 3D Gaussian beam profiles. Leading order corrections on the longitudinal electric field for the diffraction angle expansion and short pulse duration are implemented. The model featured control on parameters for focal spot position and temporal pulse center and longitudinal magnetic fields for out-of-plane laser polarization. An 800 $nm$ laser with a spot size of few $\,\mu$m was initiated at the zero density region with a normalized vector potential value of $a_0=\,$0.02. The spot size is chosen to provide suitably large field regions for the injection of the probe beam. A matched plasma channel width of 250$\, \mu$m was defined for a plasma density of 1.79$\times$10$^{22}\,$m$^{-3}$ to ensure a constant laser spot size during the propagation and a 30$\,$cm long dephasing length that is equal to the foreseen plasma length for future experimental studies. Experimentally, a cryogenically-cooled gas jet operated at many atmospheres of pressure may be used to produce the required density \cite{denseplasma}. The laser used to ionize the HIT gas is within the blowout locally. The created electrons are captured and accelerated to relativistic energies by the strong electric fields associated with the plasma wake.


\begin{figure*}[t]
    \centering
{\includegraphics[width=\textwidth]{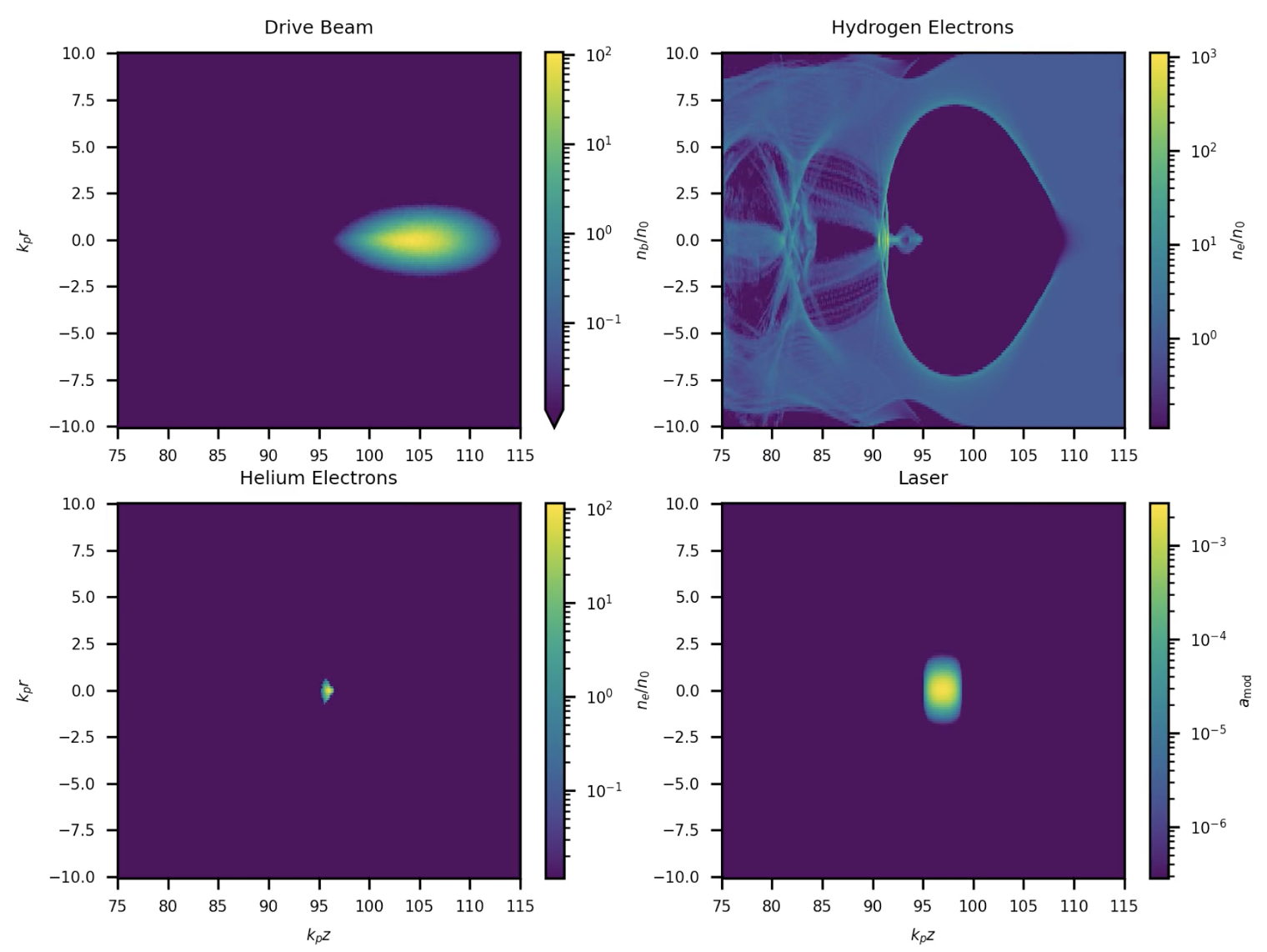}}
    \caption{Top left: Driver beam propagating through plasma shown on the top right. Bottom left: Generated witness beam using laser ionization, and the bottom right plot shows the dephasing of the laser.}  
    \label{fig:osiris_schematic}
\end{figure*}

A 1.5$\,$nC electron drive probe beam with a $\gamma=$ 20000, a matched spot size of $\sim$1$\,\mu$m and longitudinal length of 12.15$\,\mu$m. The injection phase is determined so that no accelerating field acts on the probe beam while the focusing field is larger than zero and at its maximum value. The transverse ($E_r$ and $B_{\theta}$) and longitudinal wakefields ($E_x$), are related to each other through Panofsky-Wenzel theorem \cite{Esarey_2009}. The experiment planned at FACET-II builds off the first successful demonstration of plasma photocathode at FACET \cite{Deng2019}. The planned experiment will use an ionization laser pulse injected collinearly rather than perpendicular to the direction of beam propagation to generate ultracold electron beams.

There are unique challenges involved with simulating betatron radiation from this experiment. Simulating the beam ionization is beyond the capabilities of the Quasi-static PIC code used in model \ref{sec:model2}, and achieving the required resolution using EPOCH, required computational resources well exceeding our capabilities. In this experiment, the obtainable normalized emittance for the witness beam is at the single ${(\mu rad)}$ scale. The plasma photo-cathode release a laser pulse which releases Helium electrons and forms the trapped witness bunch. Plasma photocathodes generated ultrabright electron beams can drive X-ray free electron lasers close to the cold beam limit to produce coherent X-ray pulses of attosecond-Angstrom \cite{Habib2023}.

In contrast, the oscillating fields of the laser pulse do ionize both LIT and HIT media already at intensities orders of magnitude below the intensities needed to drive plasma waves. Here, the Ti: Sapphire release laser pulse with a duration is collinear with the electron bunch driver propagation axis and follows the electron bunch traveling to the right at a distance of approximately $\mu m$ traveling to the right. A Trojan horse stage can act as a beam quality and brightness transformer. The incoming driver bunch is used to produce a witness bunch with a much higher brightness than the driver itself. Generally, there will always be a trade-off between bunch charge and emittance, and the optimum compromise may vary for different applications. Also, betatron radiation schemes are great since it is easy to produce and control betatron oscillations by releasing the HIT electrons off-axis. One should aim to optimize bunch compactness, charge, and emittance, while the energy spread of the electron bunch driver is of almost negligible importance. The Trojan horse scheme promises to generate tunable electron bunches with dramatically decreased emittance and increased brightness. Many diagnostic systems needed for characterizing the beam will be available at FACET-II. These include the betatron radiation spectrum via a Compton and pair spectrometer; the downstream beam imaging systems to determine phase space dilution of accelerated beams; and momentum-resolving spectrometers.

Angular spread diagnostics are useful ones that can separate the drive and witness spectra. The radiation analysis for the Trojan horse will be complicated because the generated witness beam will be low energy, and the radiation generated will be in a few keV ranges. However, this will not be an issue once the radiation is generated in the Trojan horse experiment. We can put detectors in a vacuum and outcouple the radiation using a spectrometer. Coherent synchrotron radiation will still be an issue, but we can use chicanes to minimize the effect. There will be a high correlation for the PWFA case. Adding compatibility with a new code is relatively simple. There are also disadvantages, as computational overhead is higher than more tightly integrated code. Tracking on multiple nodes, resuming from a progress file, and polar angular grid are also implemented to calculate the particle trajectories. 

For the most accurate and computationally intensive model, we used the 3D PIC code \texttt{OSIRIS} \cite{osiris} to generate particle trajectories which we input into the same LW code used with QuickPIC. Sampling a subset of particles from \texttt{OSIRIS} was more complicated than from \texttt{QuickPIC} because the former has variable-weighted particles. At the same time, the latter uses uniform particle weights, and the version of \texttt{OSIRIS} does not have built-in sampling functionality. In order to correctly sample, an initial simulation was performed where all the particles and weights used in the simulation were dumped, but very few output files were generated to prevent disk write bottlenecks and the generation of enormous data files. Next, particles are sampled with replacement, where the probability of sampling particle $i$ is given by $p_i = w_i / \sum_j w_j$ where $w_i$ is the weight of the $i$-th particle. After particles were sampled, redundantly sampled particles were combined by adding their weights. After this, a second simulation is run where the input file \texttt{OSIRIS} is instructed to only output particles with IDs in the list of sampled particles. The trajectories are computed from the output files, and the LW integral is computed using Eq.~(\ref{eq:lw1}) and Eq.~(\ref{eq:sumrad}) where the $\bm{V}_i$ are scaled by the square root of the particle weights.

\begin{table}
\caption{\label{tab: E-310_Trojan_Horse} Laser and electron beam parameters for Trojan horse experiment at FACET-II. }
\begin{ruledtabular}
\begin{tabular}{cc}
Parameter & Value \\
        \hline
        Species & H  \\
         Laser wavelength ${(nm)}$  & 800  \\
         Tau ${(fs)}$  & 50  \\
         Laser $a_0$  & $0.02$  \\
         Plasma wavelength ${(\mu m)}$& 250  \\
         $nH2(LIT) = nHe(HIT)$ ${(cm^{-3})}$ & 1.789e16 \\
         $n_0$ ${(cm^{-3})}$& 1.79e16 \\
          $\omega_p$ ${(\mu m)}$ & 100  \\
          $k_p^{-1}$ ${(\mu m)}$& 39.79  \\
          Beam peak density & 9.3e23= 52  \\ 
        \hline
        Drive beam parameter & Unit\\
        \hline
        $E$ ${(GeV)}$ & 10  \\
        $Q$ ${(nC)}$& -1.5  \\
        $Q tilda$ & $8.3$ \\
        $\Omega_l$ & $313$   \\
        $N$ & $3.1 \times 10^{9}$\\
        Laser beam waist ${(\mu m)}$ &7  \\
        $\sigma_x \mathrm{unmatched}$ ${(\mu m)}$& 4.5  \\
        $\sigma_y \mathrm{unmatched} $ ${(\mu m)}$& 4.5  \\
        $\sigma_z$ ${(\mu m)}$ & 12.15  \\
        $\epsilon_{n,x}$ ${(\mu m)}$ & 5  \\ 
        $\epsilon_{n,y}$ ${(\mu m)}$ & 5  \\
      \hline  
\end{tabular}
\end{ruledtabular}
\end{table}

\begin{figure}[h!]
    \centering  \includegraphics[width=\columnwidth]{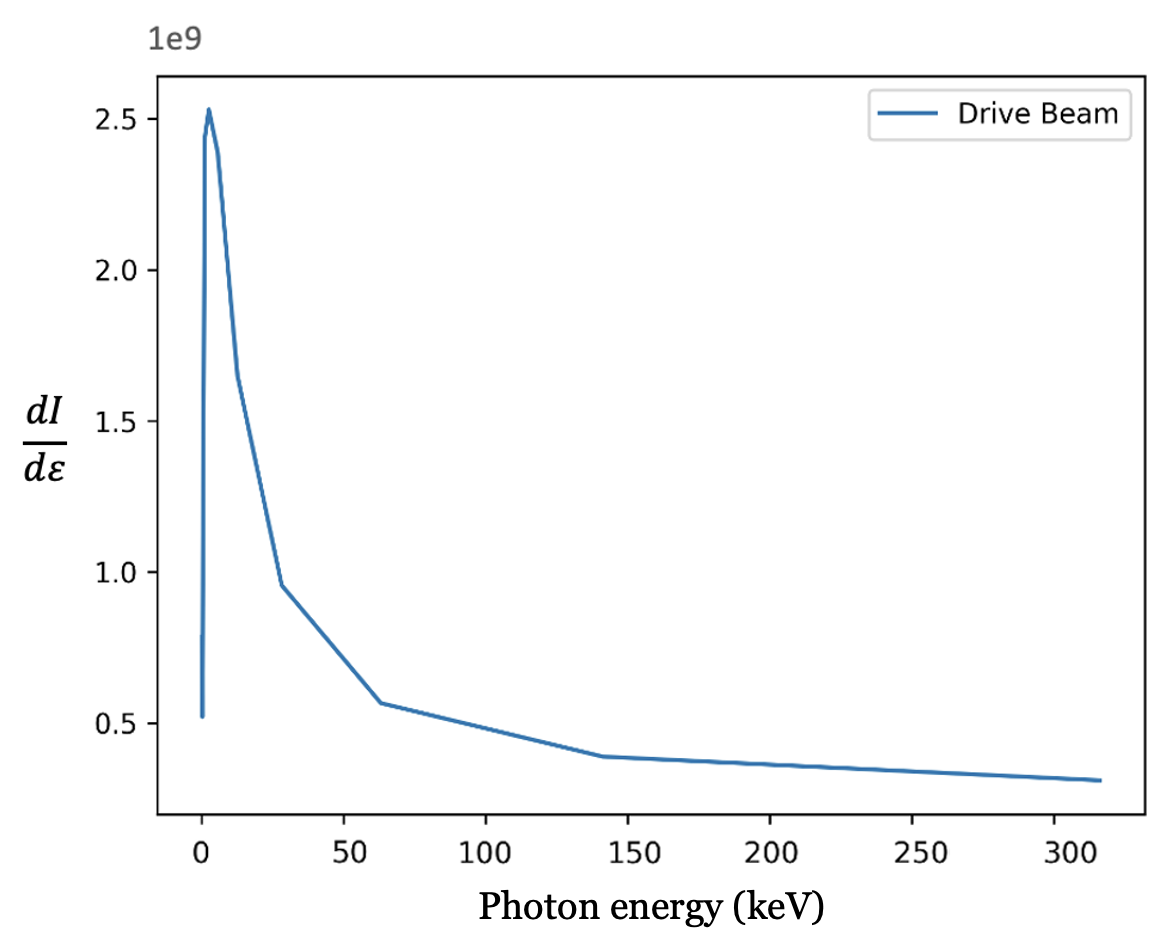}
    \caption{Photon energy spectrum of the radiation emitted by the driver bunch computed using \texttt{OSIRIS} and LW code.}
    \label{fig:sxanalytic1}
\end{figure}
\begin{figure}[h!]
    \centering
    \includegraphics[width=\columnwidth]{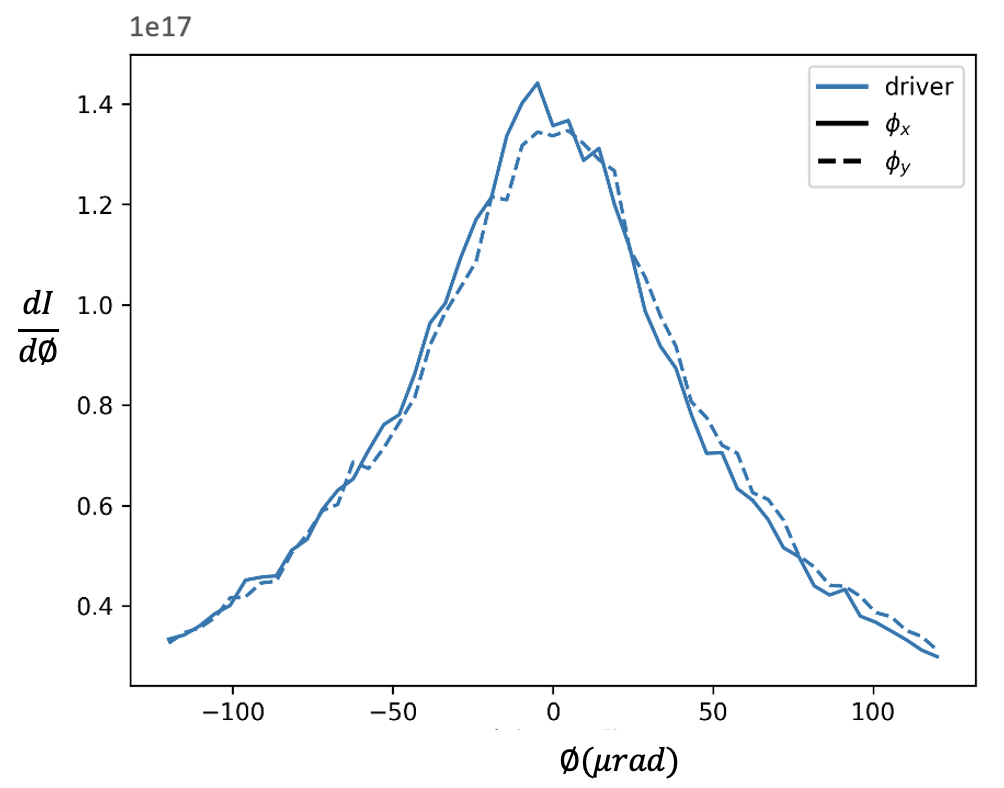}
    \caption{ 1-D angular $\phi_x$ and $\phi_y$ distribution of betatron radiation generated by driver beam.}
    \label{fig:sxanalytic2}
\end{figure}

A plasma profile comprising a vacuum section followed by an increasing short ramp section was implemented. One of the experimental goals of FACET-II is the demonstration of the creation of high brightness beams from a 'Trojan horse' plasma photocathode \cite{trojanhorse, trojan_horse_2018}. This is a scheme in which a beam is created in a plasma wake through laser ionization of neutral gas. In this scheme, the target comprises both a high ionization threshold (HIT) gas and a low ionization threshold (LIT) gas. The LIT gas is preionized, and the drive beam creates a strong plasma wave blowout. \texttt{OSIRIS} does not output tags so we modified the \texttt{OSIRIS} code. Sampling is done with replacement and redundantly sampled macro particles combined. After modification in the code the particle tags were output to generate the trajectory data, and associated the trajectories with weights.



In Fig. \ref{fig:sxanalytic1} and Fig. \ref{fig:sxanalytic2}, synchrotron radiation spectra for the driver beam are shown, and the spectrum is narrow. There are unique challenges involved with simulating betatron radiation from this experiment. Simulating the beam ionization is beyond the capabilities of the Quasi-static PIC code used in model \ref{sec:model2}, and achieving the required resolution using required computational resources well exceeding our capabilities. In recently published paper \cite{osiris_radio}, algorithm characterize electromagnetic waves by implementing the LW potentials to extract radiation emission. We compared our model with OSIRIS Radio code and results are similar.

\subsection{Benchmark using full Particle-in-Cell EPOCH with Monte Carlo QED radiation} \label{EPOCH}

Despite the sophistication of the PIC and LW radiation models, some effects still require a higher level of sophistication to simulate. Monte Carlo QED radiation models treat high-energy photons as discrete particles, and electrons have some probability of emitting them. Statistical properties of the photons could matter, especially at extremely high energies, quantum recoil, statistical properties, and strong field effects. An example is $\texttt{EPOCH}$, a 3D fully explicit PIC code that uses a Monte Carlo QED model to simulate radiation generation \cite{Epoch_arber_2015,epoch_Ridgers2014}.

We found that accurately simulating radiation using this method exceeded our computational resources. A primary challenge with 3D explicit codes is the artificial slow-down of the speed of light on a finite difference time domain (FDTD) grid. This means, for example, that a relativistic electron propagating along a straight line with constant velocity in free space will nonphysically emit numerical Cherenkov radiation (NCR) at wavelengths corresponding to the grid cell size and may even grow as an instability by imprinting into the current profile \cite{Num_cherenkov}. In EPOCH, we use a dispersion-reduced FDTD solver \cite{lehe} and an 8-point, compensated \cite{J_Vay} linear current filter to mitigate this effect. Such schemes are imperfect and can slightly alter the Fourier content of fields at the grid resolution. However, the radiation model in EPOCH is photon-based, not field-based, and the emitted radiation wavelengths are well beyond the grid resolution. Therefore, we expect minimal interference from the smoothing filter and microscopic details of the dispersion on our results while retaining the benefit of smooth fields to be used in QED calculations. One of the primary challenges associated with using EPOCH is that extensive computational resources are needed to correctly resolve physically relevant length scales, especially in the matched beam case when the beam spot size is small. The domain is set up with 512$\times$512$\times$512 cells per in the longitudinal and transverse directions, allowing the minor features in those directions to be resolved. Drive and witness beams in the plasma are represented with macro-particles per cell, assuming an immobile neutralizing background. 

\begin{figure}[h!]
    \centering   \includegraphics[width=\columnwidth]{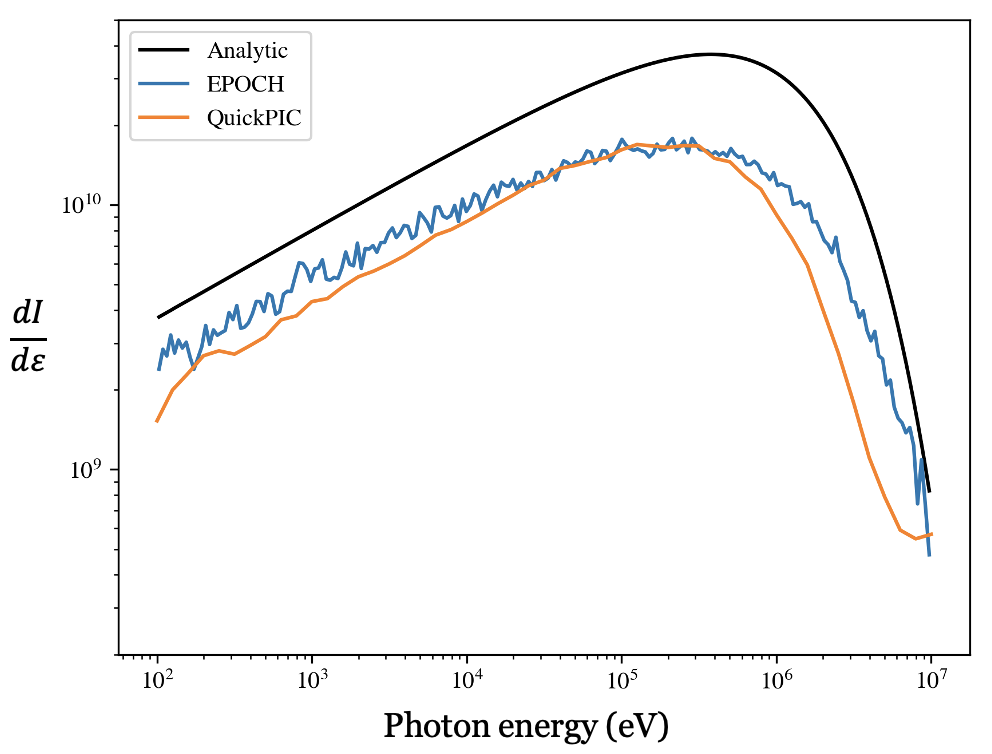}
    \caption{Benchmark of photon energy spectrum of the radiation emitted by the driver bunch computed using \texttt{EPOCH} code, analytic and \texttt{QuickPIC} code. In the analytic plot the energy loss is the drive beam is not considered.}
    \label{fig:epoch_1}
\end{figure}

\texttt{EPOCH} is the most sophisticated and computationally intensive betatron radiation model. In EPOCH input files, a smoothing function is applied to the current generated during the particle push. It helps to reduce noise and self-heating in a simulation. It can be substantially tuned to damp high frequencies in the currents and can be used to reduce the effect of NCR. Once we turn on the current filtering, we can set the following keys: smooth iterations, integer number of iterations of the smoothing function to be performed. If not present defaults to one iteration, more iterations will produce smoother results but will be slower.
In Fig. \ref{fig:epoch_1} benchmark of photon energy spectrum of the radiation emitted by the driver bunch computed using \texttt{EPOCH} code, analytic and \texttt{QuickPIC} code. The analytical curve is much higher than the \texttt{EPOCH} and \texttt{QuickPIC} as the
analytic plot does not consider the energy loss is the drive beam. However this would not be an issue in the case of witness beam which is more stable or beams with zero emittances.

\begin{figure}[h!]
    \centering   \includegraphics[width=\columnwidth]{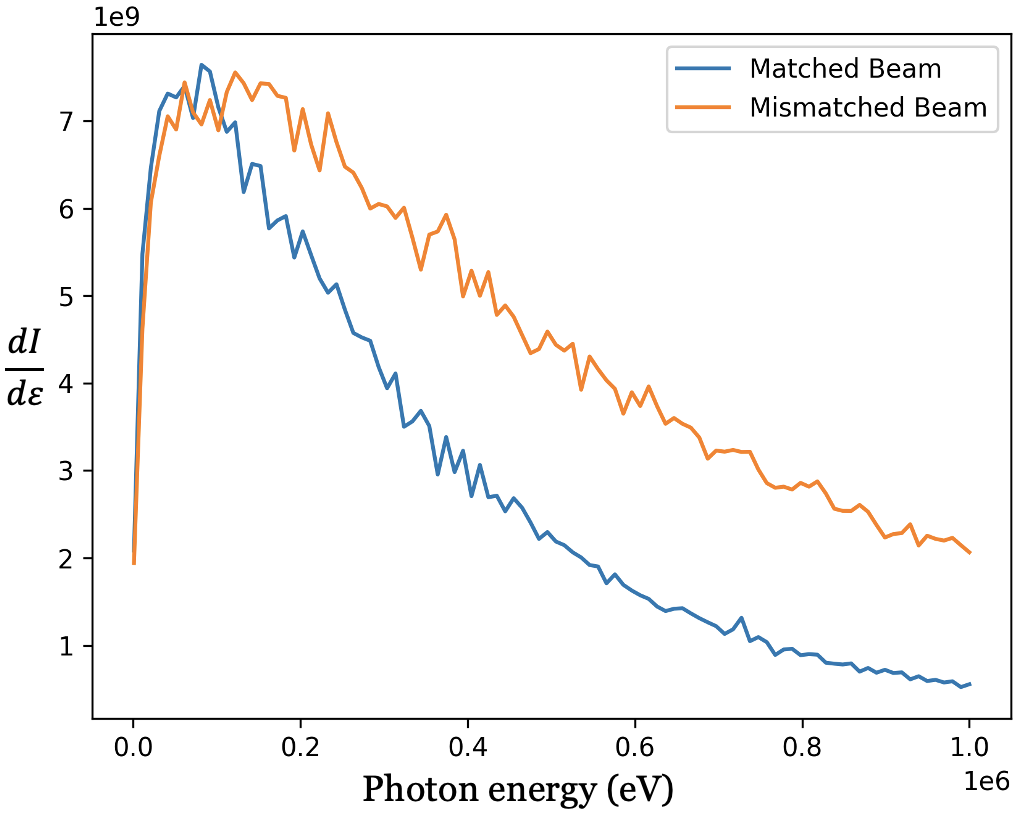}
    \caption{Photon energy spectrum of the radiation emitted by the driver bunch computed using \texttt{EPOCH} code for matched and mismatched beam at the plasma entrance. }
    \label{fig:epoch_2}
\end{figure}

Gamma rays emitted from the betatron radiation process in underdense plasma wakefields and high field-laser-induced Compton scattering produce unique experimental signatures which can reveal interaction physics at the challenging ultra-short spatial and temporal experimental scales. Additionally, radiative diagnostics provide a non-intercepting, non-destructive probe of the extreme high-field environments relevant to advanced acceleration techniques. An electron beam is matched to the plasma density if the beam envelope experiences no oscillations. In Fig. \ref{fig:epoch_2}, the emitted radiations from a matched and mismatched beams are plotted in orange and blue colour, respectively. Higher photon energy is achieved in the unmatched case because of the high oscillation amplitudes of the electrons. 

\section{\label{sec:pwfa} FACET-II PWFA simulations}

This section introduces a parameter set for a PWFA at FACET-II. The simulation parameters were summarized in Table \ref{tab:model2validation1params}. We characterized the radiation for the parameter set based on what is feasible for PWFA experiments at FACET-II. We demonstrated the effect caused by focusing and defocusing at betatron periods. Scalloping in the driver beam's head results in deterioration of the beam after a few centimeters, propagation in plasma.

\begin{figure*}[]
    \centering
   \subfloat[]{ \includegraphics[width=0.32\textwidth]{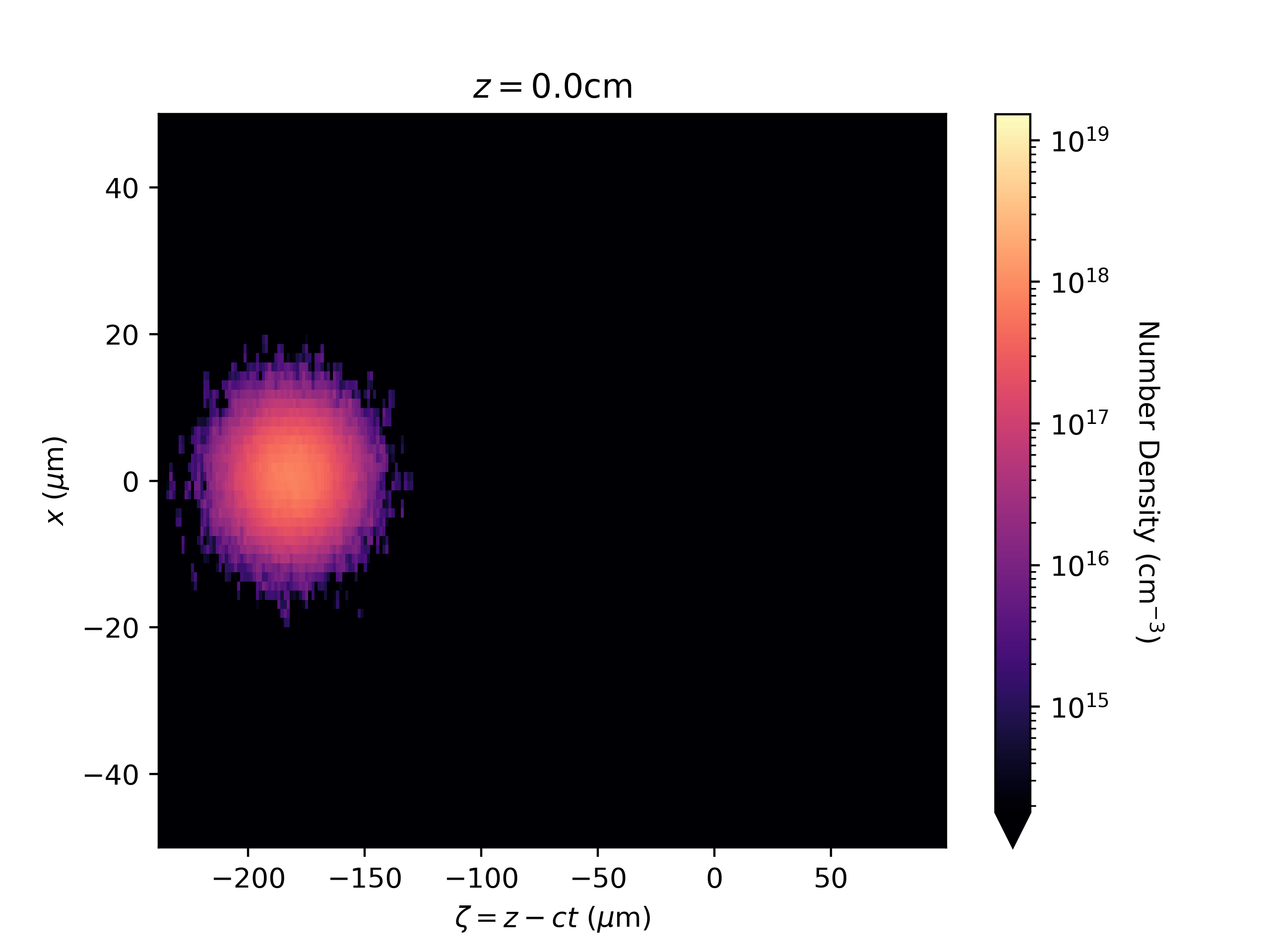}}
    \subfloat[]{\includegraphics[width=0.32\textwidth]{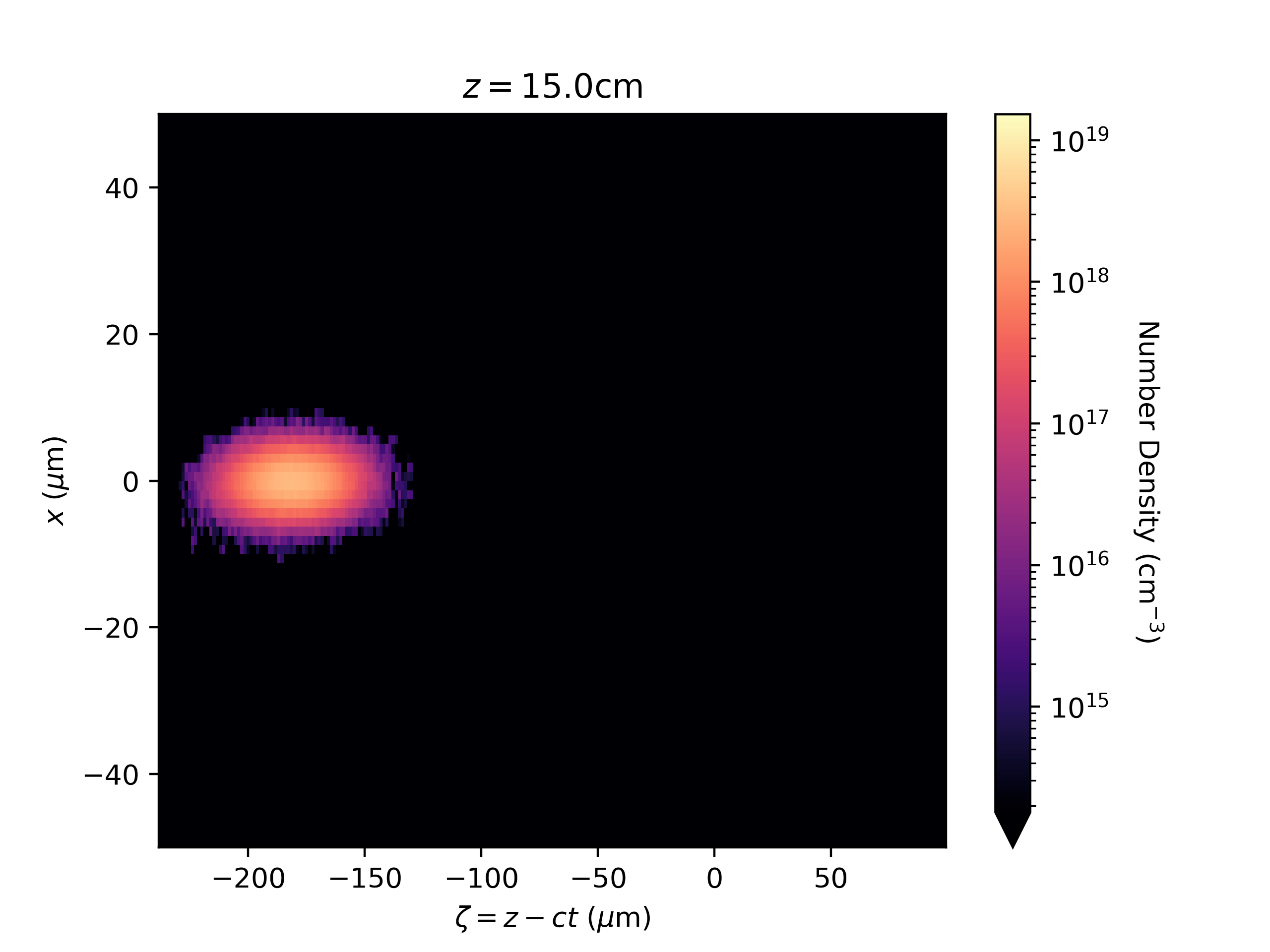}}
   \subfloat[]{ \includegraphics[width=0.32\textwidth]{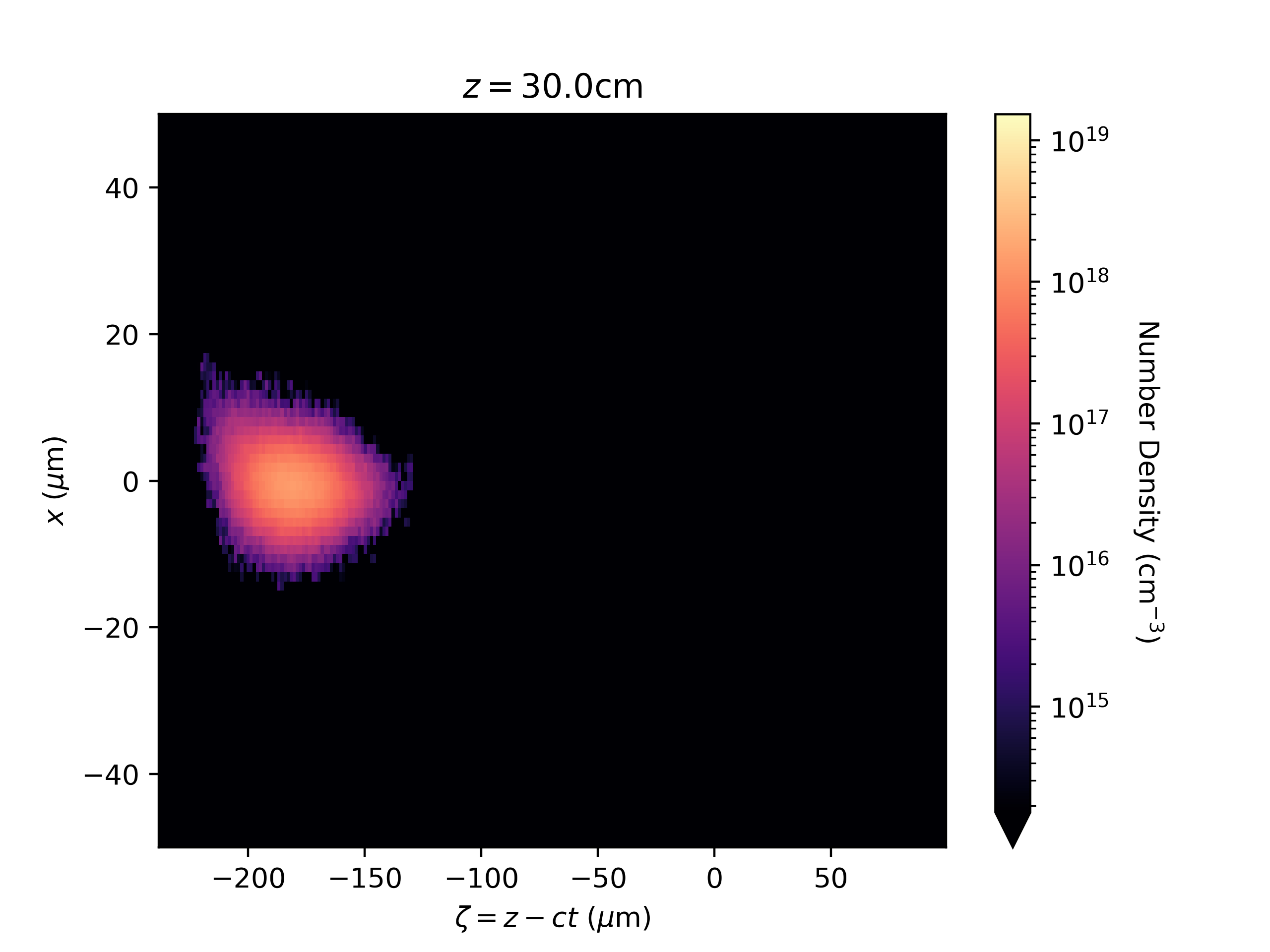}}
   \caption{ Beam electron number density x-z \textbf{(a)} slices at $z = 0 \, \mathrm{cm}$, \textbf{(b)} slices at $z = 15 \, \mathrm{cm}$, \textbf{(c)} slices at $z = 30 \, \mathrm{cm}$. }
    \label{fig:prototype_witness_1}
\end{figure*}

\begin{figure*}[]
    \centering
   \subfloat[]{ \includegraphics[width=0.32\textwidth]{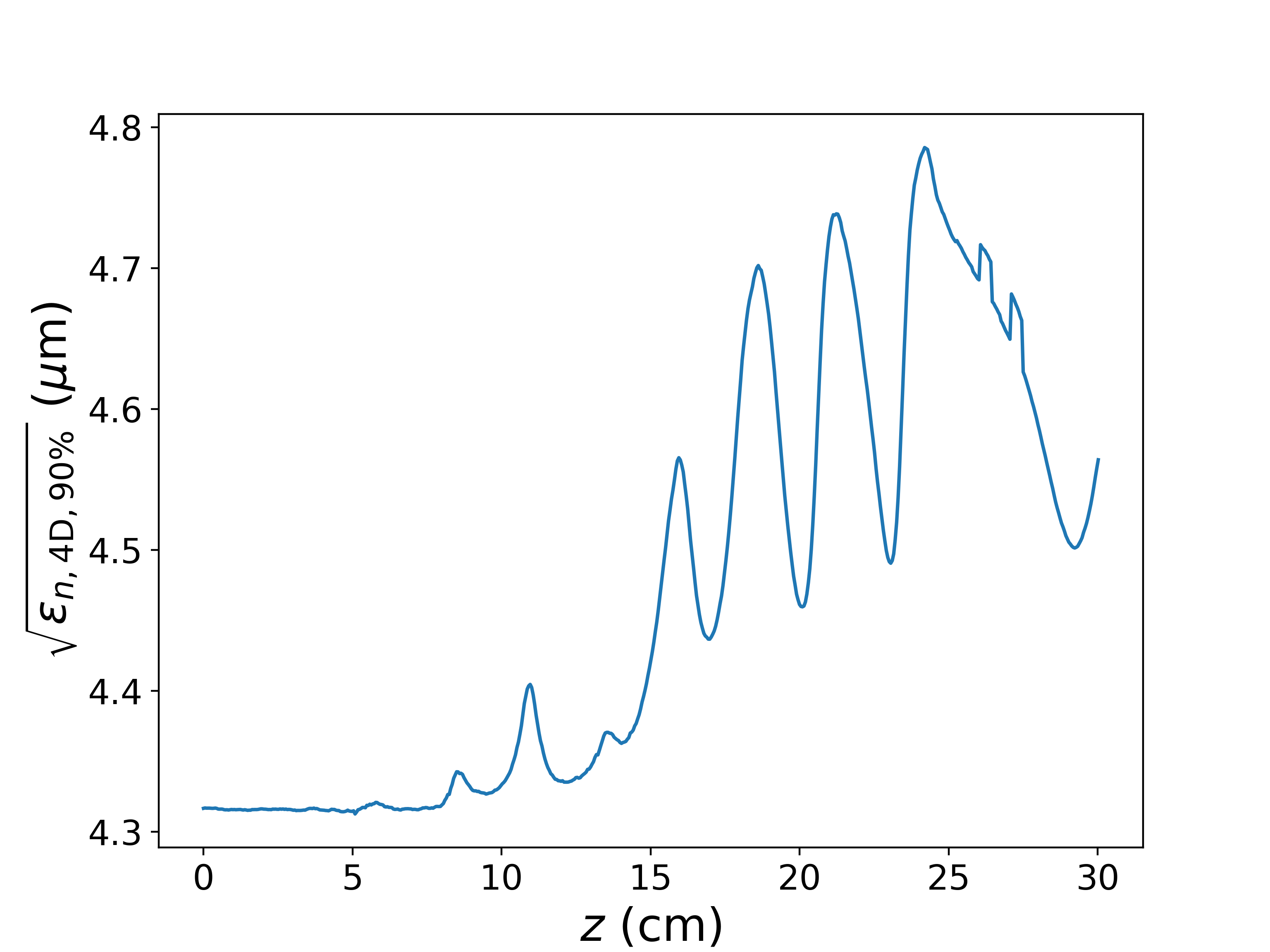}}
   \subfloat[]{ \includegraphics[width=0.32\textwidth]{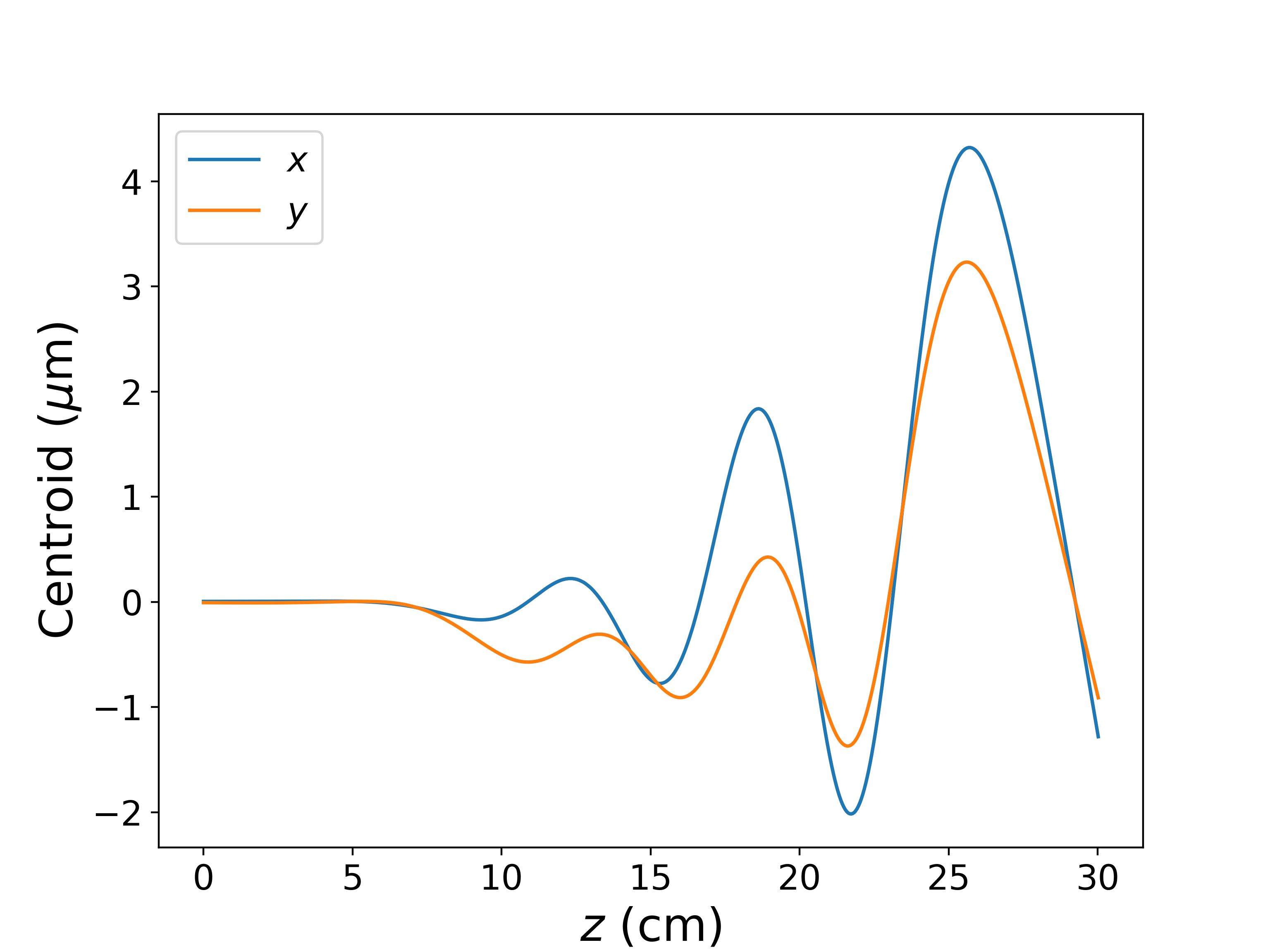}}
   \subfloat[]{ \includegraphics[width=0.32\textwidth]{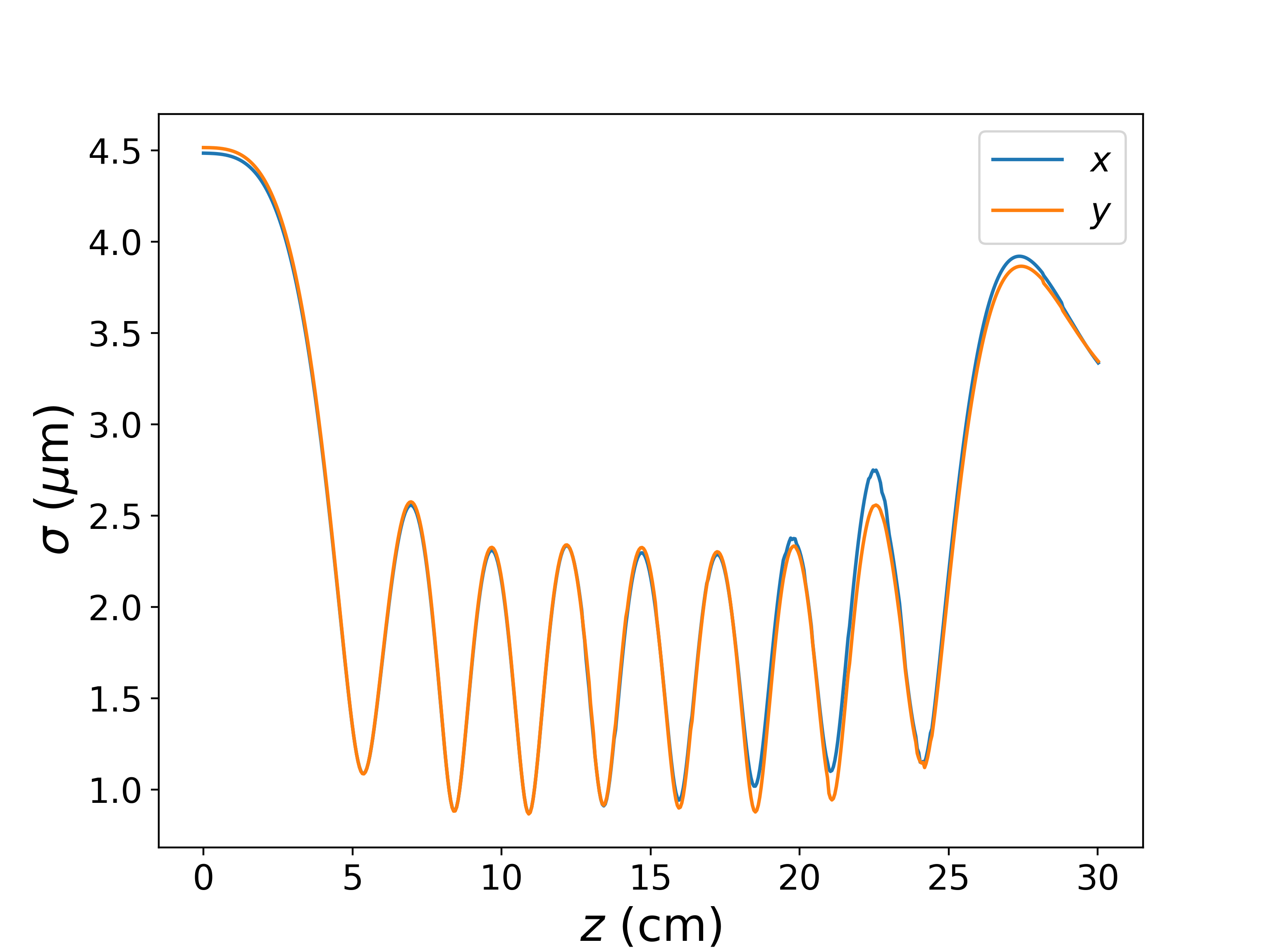}}
   \caption{\textbf{(a)} Witness beam emittance growth over a long propagation distance, \textbf{(b)} witness beam centroid and \textbf{(c)} is the witness sigma evolution.}
    \label{fig:prototype_witness_2}
\end{figure*}

\begin{figure*}[]
    \centering 
   \subfloat[]{ \includegraphics[width=0.32\textwidth]{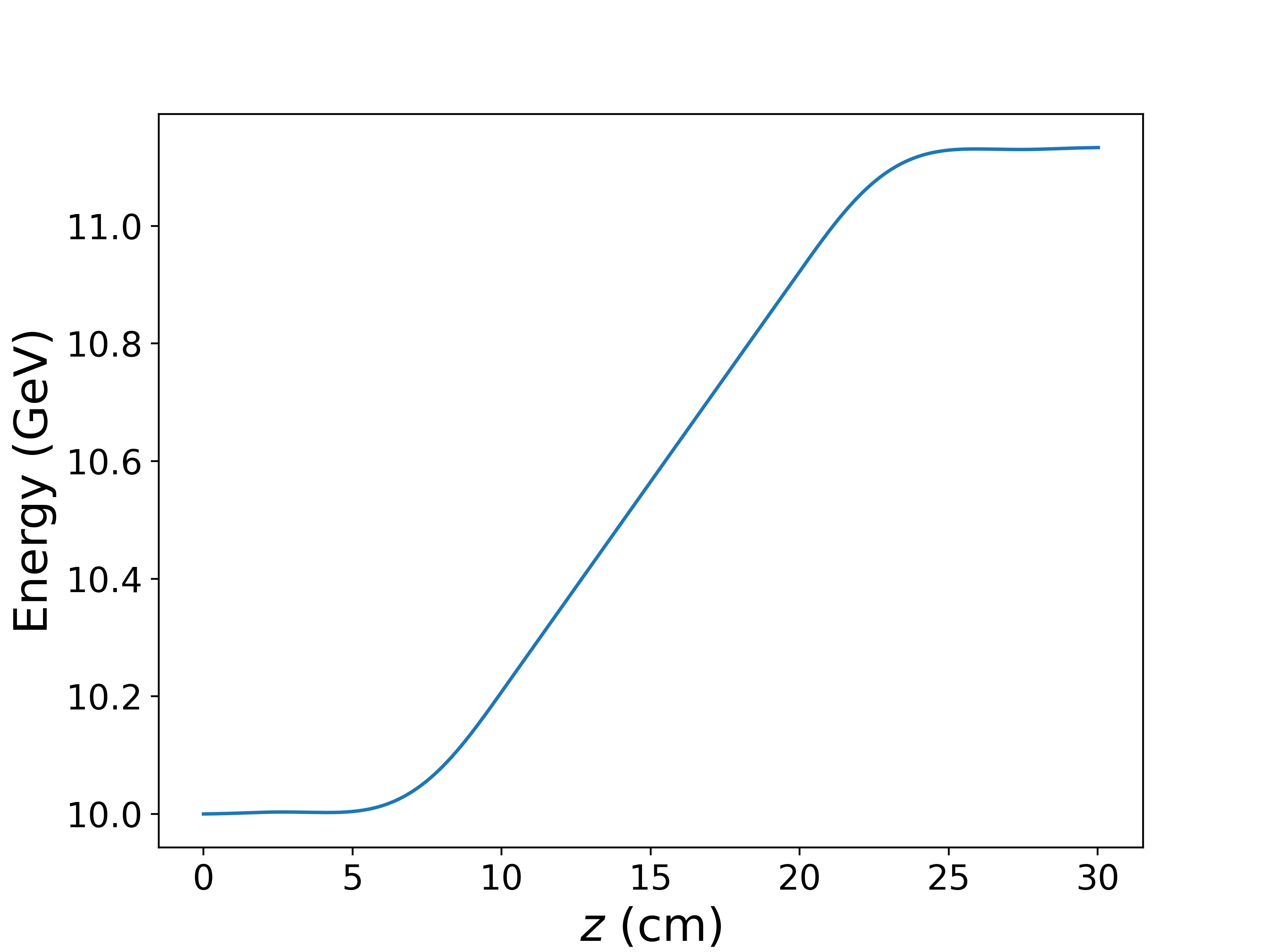}}
   \subfloat[]{ \includegraphics[width=0.32\textwidth]{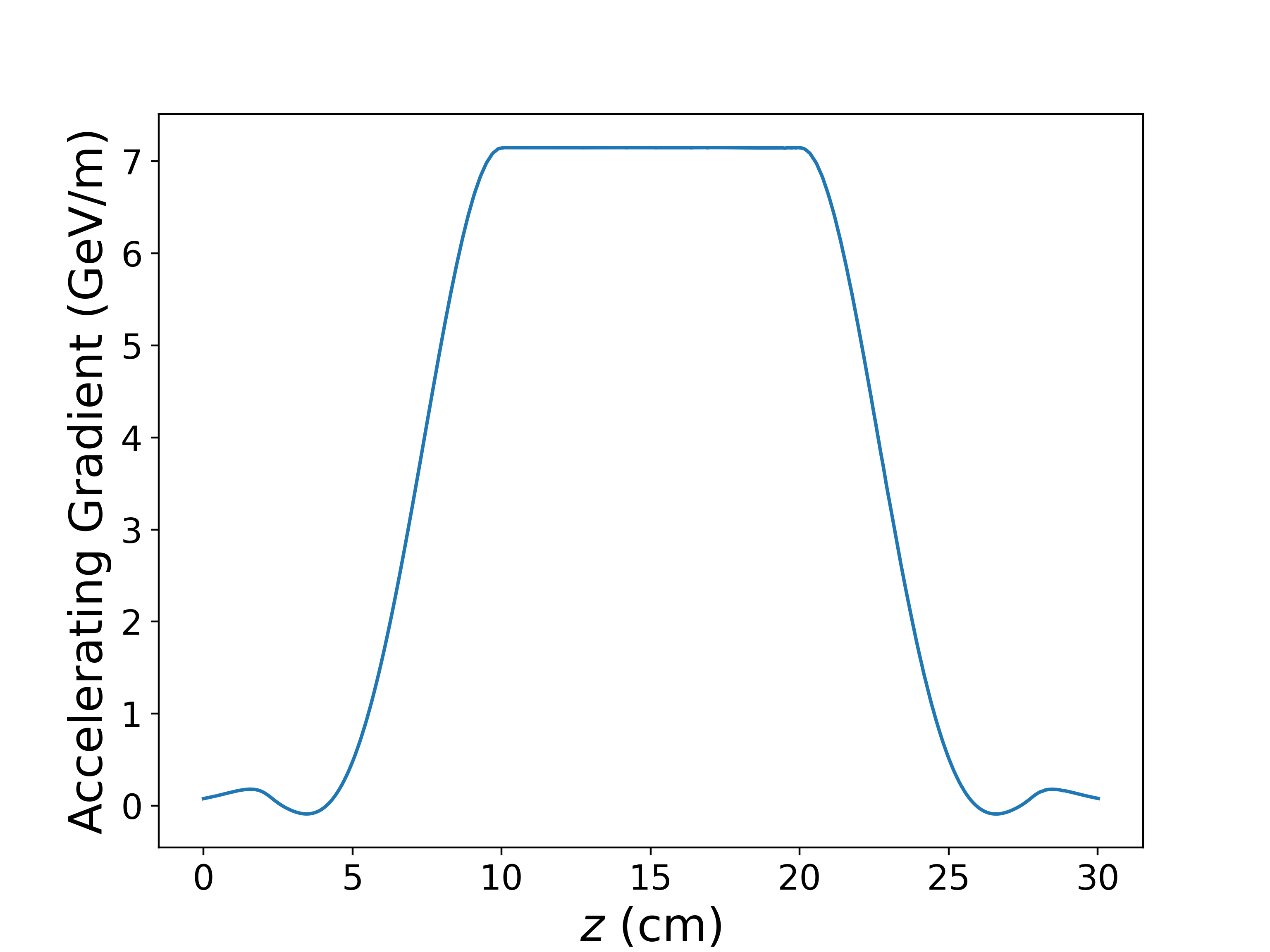}}
   \subfloat[]{ \includegraphics[width=0.32\textwidth]{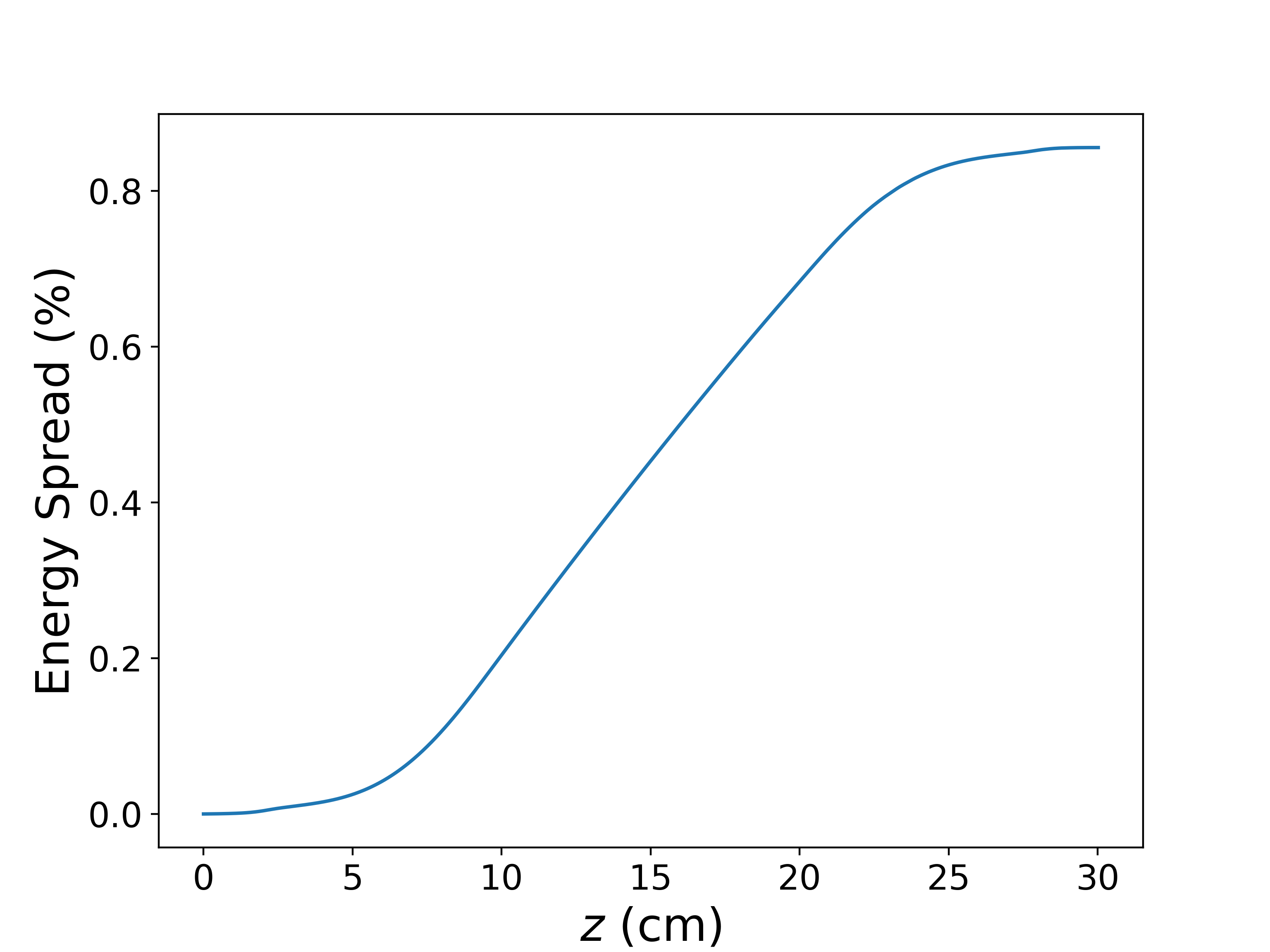}}
      \caption{\textbf{(a)} Energy gain in witness beam, \textbf{(b)} accelerating gradient over a ramp plasma profile and \textbf{(c)} energy spread in the witness beam.}
    \label{fig:prototype_witness_3}
\end{figure*}

\begin{figure*}[]
    \centering
    \subfloat[]{\includegraphics[width=0.47\textwidth]
    {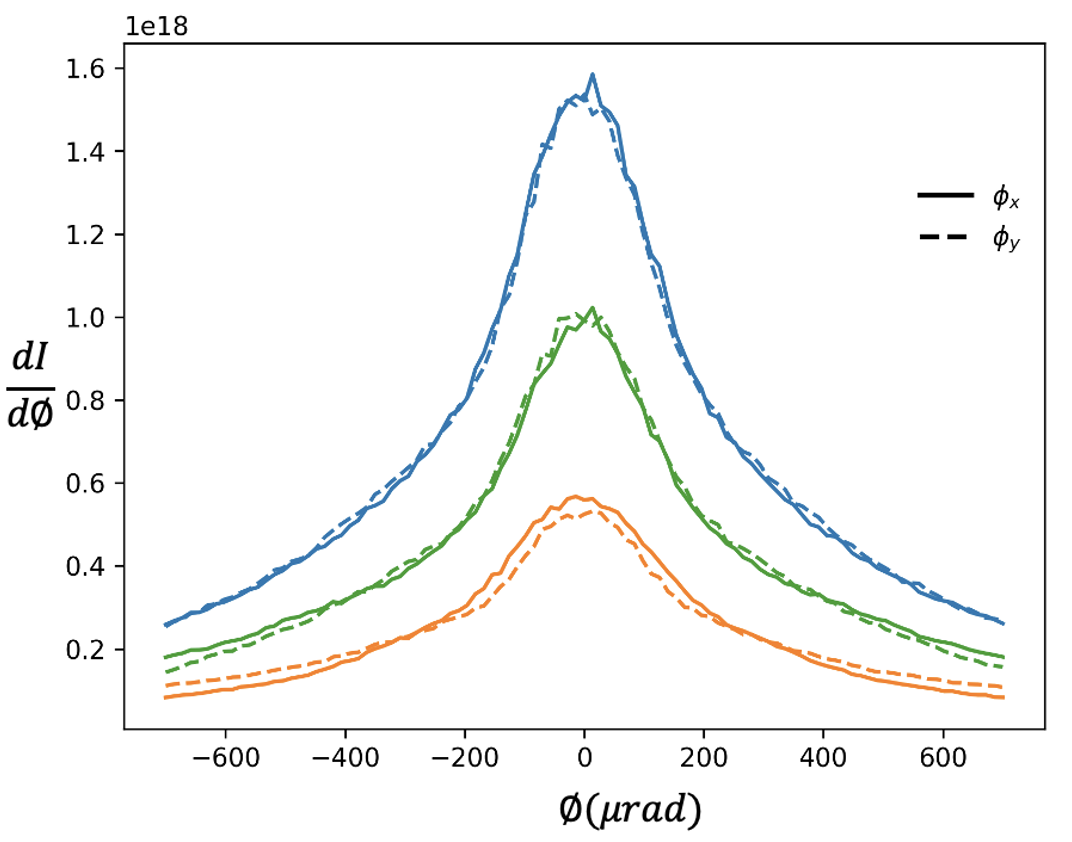}}
   \subfloat[]{ \includegraphics[width=0.47\textwidth]{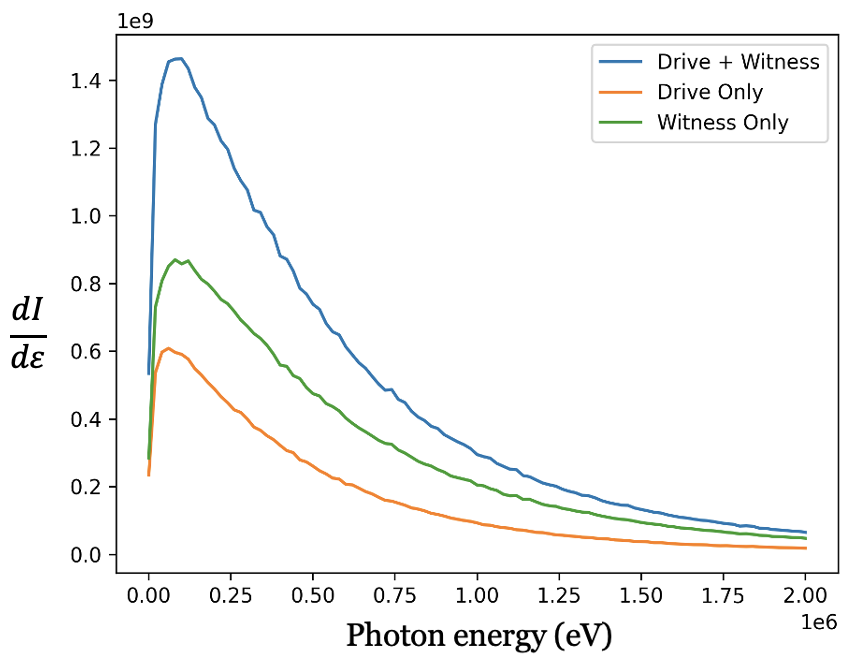}}
    \caption{\textbf{(a)} 1-D angular distribution of betatron radiation generated by driver and witness beam \textbf{(b)} betatron radiation spectrum from the prototype simulation for both the beams charge is 1.5 nC.}
    \label{fig:prototype_radiation}
\end{figure*}

We are simulating two bunch scenario of drive and witness in PWFA. The initial emittance of the drive beam was 3.2 $\mu m$ and 3 $\mu m$ in the $x$ and $y$ directions, respectively. However, the energy spread in the driver beam is significantly less; hence the emittance is not growing fast. 
The centroid oscillations in $x$ and $y$ are different because the emittance is different in both directions. 
The drive beam starts at a spot size of 5 $\mu m$ and focuses down to 1 $\mu m$ because of the linear forces. The initial energy of the driver beam is 10 GeV, after propagating for a distance of $z = 30 \, \mathrm{cm}$, the beam loses approximately 0.8 GeV energy to the plasma electrons. As the ramp profile is semi-Gaussian, the driver beam decelerates quickly at lower plasma densities and maintains a 4 GeV/m decelerating gradient for a uniform plasma density. 


In Fig. \ref{fig:prototype_witness_1} witness beam electron number density x-z slices at (a) $z = 0 \, \mathrm{cm}$, (b) $z = 15 \, \mathrm{cm}$, (c) $z = 30 \, \mathrm{cm}$ are shown. The witness beam is more stable than the driver beam because it is in the pure ionic column. The witness beam oscillates at a betatron frequency, and the oscillations will lead to betatron radiation. In Fig. \ref{fig:prototype_witness_2}{(a)} witness beam emittance growth over a longer propagation distance is plotted. 
We notice that the witness beams $90 \%$ emittance grows very little. 
The growth at longer propagation is much less than the driver beam because the witness beam is more stable and not deteriorating like the driver. 
In Fig.~\ref{fig:prototype_witness_2}{(b)} the witness beam centroid is plotted. The centroid oscillation amplitude change in the $x$ and $y$ directions is the same. In \ref{fig:prototype_witness_2}{(c)} witness sigma evolution is shown; the beam starts at 4.5 $\mu m$ and focuses down to 0.8 $\mu m$. The beam oscillates at a betatron frequency, where it focuses and defocuses. 

In Fig. \ref{fig:prototype_witness_3}{(a)} the energy gain in the witness beam is plotted for a length at the entrance of plasma (z=0), $\mathrm{cm}$, the beam energy increase to more than 11 GeV. Figure \ref{fig:prototype_witness_3}{(b)} shows that the accelerating gradient over a ramp plasma profile is 7 GeV/m. The witness beam's accelerating gradient is much less on the up and down ramp when the plasma densities are smaller, but it gradually increases with the plasma density. In Fig. \ref{fig:prototype_witness_3}{(c)}, we show the energy spread in the witness beam. The spectrum of radiation produced by the drive, witness, and both beams are shown in Fig. \ref{fig:prototype_radiation}. The noise in the spectrum is primarily a result of the finite $\phi_x$/$\phi_y$ step size. In this case, the energy of the drive and witness beams are 10 GeV each. The separation between the two beams is 150 $\mu m$. The maximum share of the radiation is produced by the witness beam, which helps diagnosing witness beam parameters. We notice that the high energy tail of the radiation is dominated by the witness beam.

\section{\label{sec:photocathode} Radiation produced by drive beam in Trojan horse experiment}

One of the experimental goals of FACET-II is the demonstration of high-brightness beams generated from a Trojan horse plasma photocathode \cite{trojanhorse, trojan_horse_2018}. 
In this scheme, a beam is created within a plasma wakefield, through laser ionization of neutral gas. The gaseous mixture is comprised of a low ionization threshold species for plasma wakefield generation, and a high ionization threshold species for particle generation. Another way of generating witness beam using laser and beam
radial fields overlap to liberate electrons from the tunneling ionization of the non ionized gas species is discussed in \cite{dragontail}.
In this section, we simulated a single bi-Gaussian beam traveling through a uniform plasma using the Quasi-static PIC code \texttt{QuickPIC} \cite{qpic1,qpic2} for the FACET-II E-310 Trojan horse parameters. 
The beam driver is assumed to have a bi-Gaussian density profile,

%
\begin{align}
%
    n_b(\xi, r)/n_0 = \left( n_b/n_0 \right) \, e^{-\xi^2/(2\sigma_\xi^2)} \, e^{-r^2/(2\sigma_r^2)}
    \label{nb}
\end{align}

where $\xi \equiv z - ct$ is the beam co-moving coordinate, $c$ is the speed of light in vacuum, $n_b \equiv (Q/e) / [(2\pi)^{3/2} \sigma_\xi \sigma_r^2]$ is the peak beam-density, $n_0$ is the plasma electron density, $Q$ is the beam charge and $\sigma_\xi$, $\sigma_r$ are the beam longitudinal and radial RMS sizes, respectively. 
The beam has initial kinetic energy $E_{k0}$, and energy spread $\delta E_{k0}/E_{k0} = 1\%$. 
The simulation parameters are shown in Table \ref{tab: E-310_Trojan_Horse}.





\begin{figure*}[]
    \centering
   \subfloat[]{\includegraphics[width=0.34\textwidth]{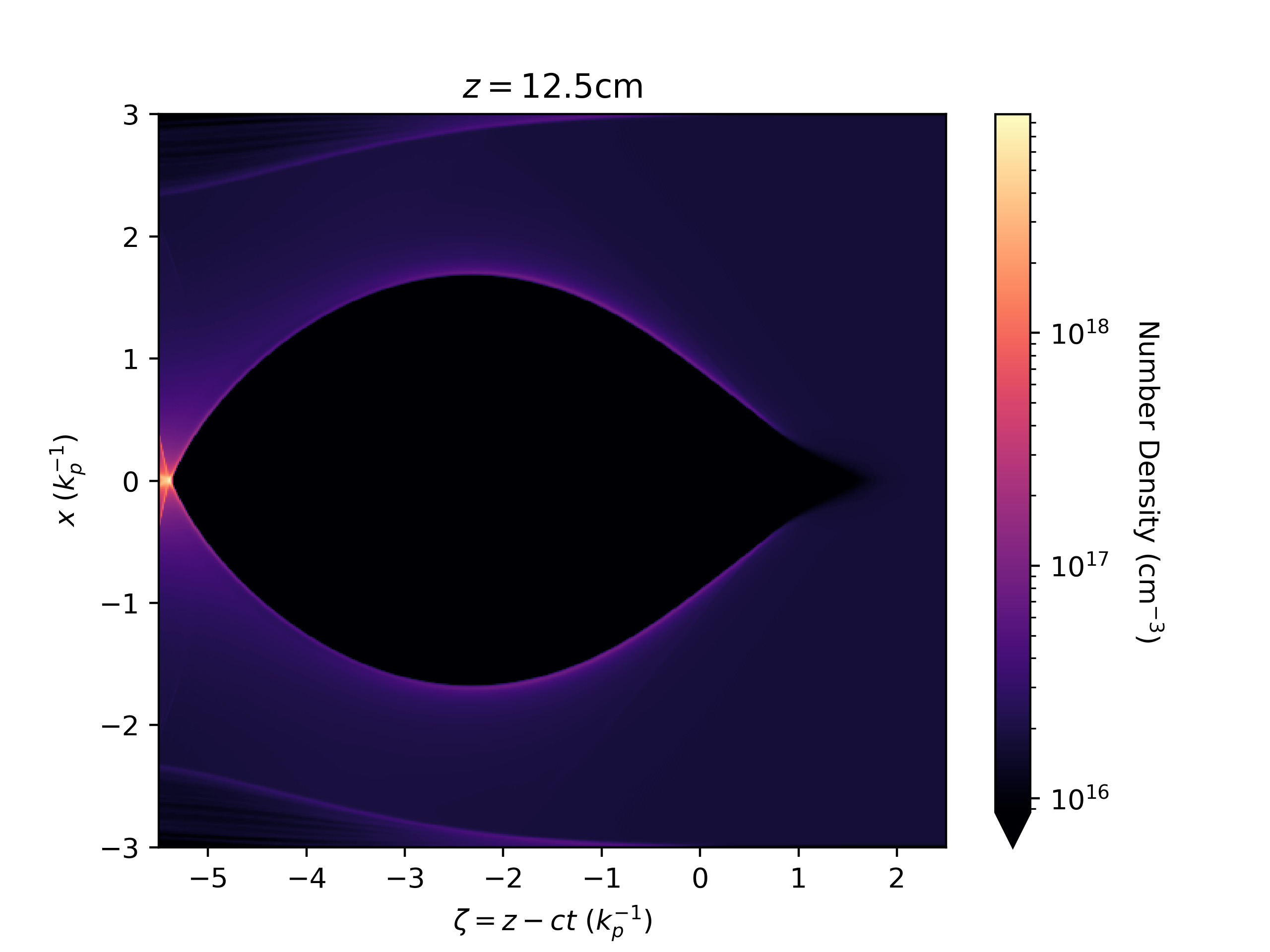}}
     \subfloat[]{ \includegraphics[width=0.30\textwidth]{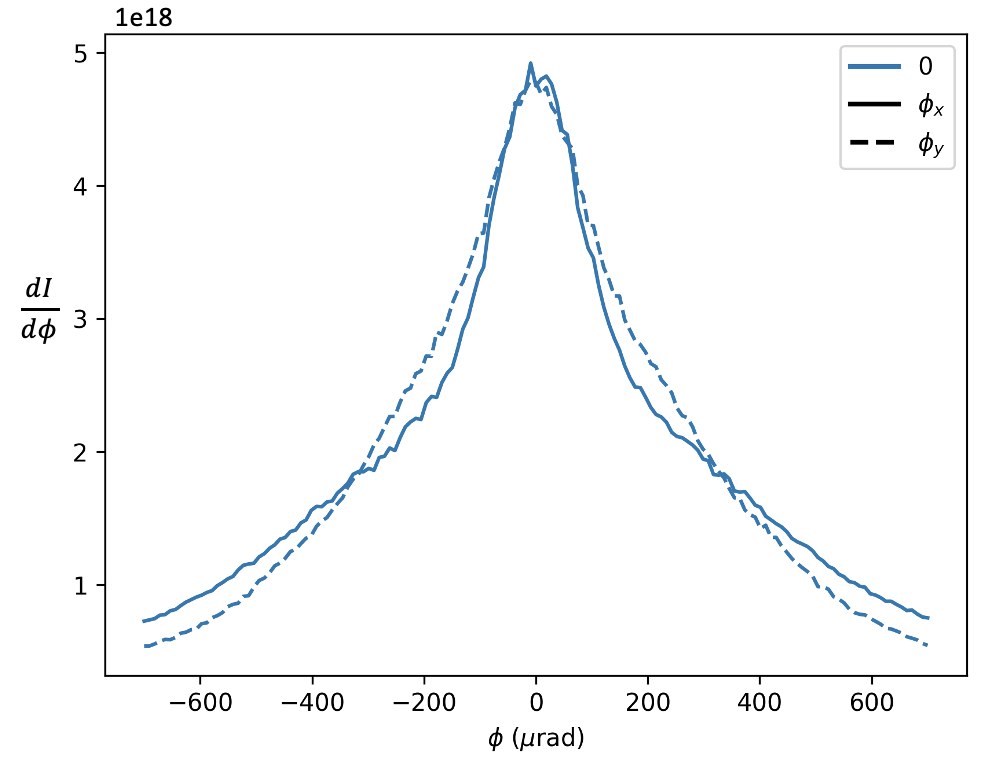}}
        \subfloat[]{\includegraphics[width=0.30\textwidth]
        {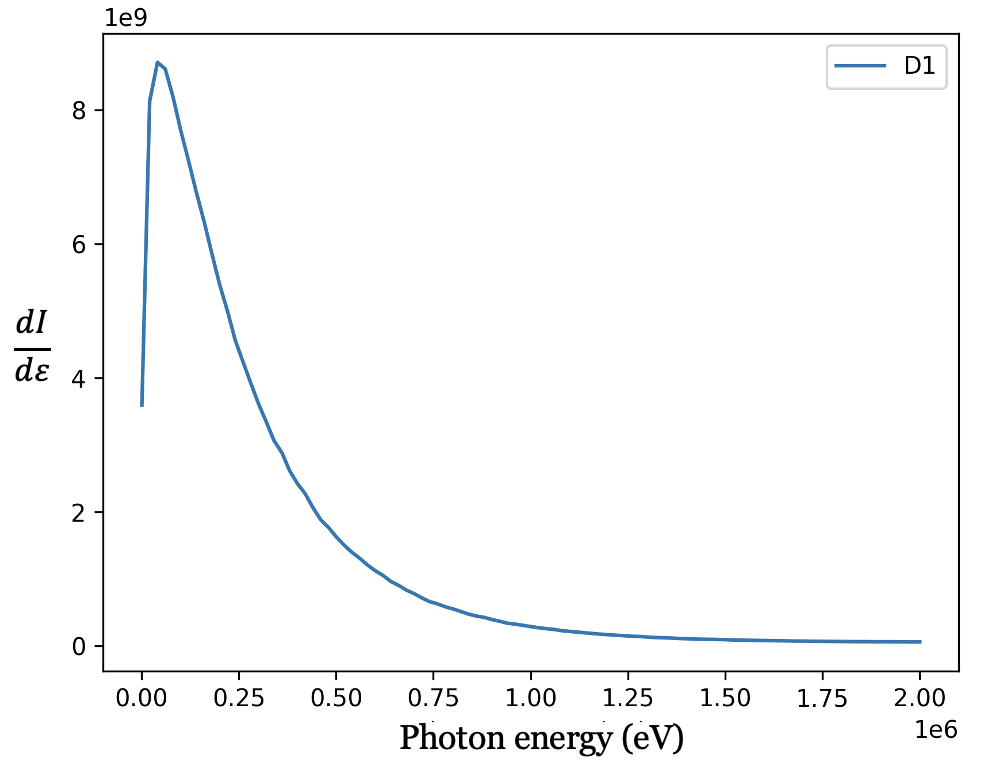}}
 \caption{(a) PIC simulation snapshots of the blowout  at $z = 12.5 \, \mathrm{cm}$, (b) 1D angular $\phi_x$ and $\phi_y$ distribution of betatron radiation generated by driver beam using QuickPIC simulations and, (c) photon energy spectrum of the radiation emitted by the driver bunch computed using QuickPIC and LW code.}
    \label{fig:drive_E310_quickpic}
\end{figure*}


Fig. \ref{fig:drive_E310_quickpic}(a) shows the PIC simulation snapshots of the blowout at $z = 12.5 \, {cm}$. The driver beam is dense enough to eject all the plasma electrons to form a strong blowout. In Fig. \ref{fig:drive_E310_quickpic}(b) 1-D angular $\phi_x$ and $\phi_y$ distribution of betatron radiation and, Fig. \ref{fig:drive_E310_quickpic}(c) photon energy spectrum of the radiation emitted by the driver bunch computed using model \ref{sec:model2}.
We measure the angular information about the
photons, critical for understanding the upstream physics, via their lateral displacement.
Further information is obtained from double differential information (angular and spectral) to assess beam dynamics and constrain modeling of plasma beam dynamics, in Trojan horse injection experiment.


\begin{figure}
    \centering
    \subfloat[] {\includegraphics[width=\columnwidth]{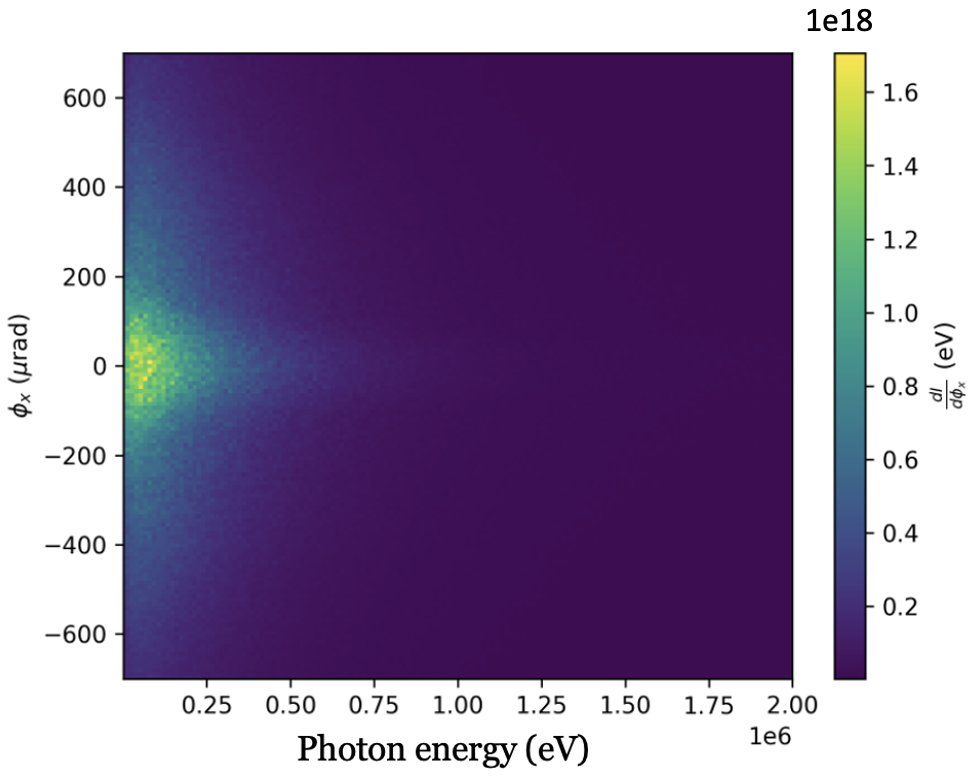}}\\
     \subfloat[] {\includegraphics[width=\columnwidth]{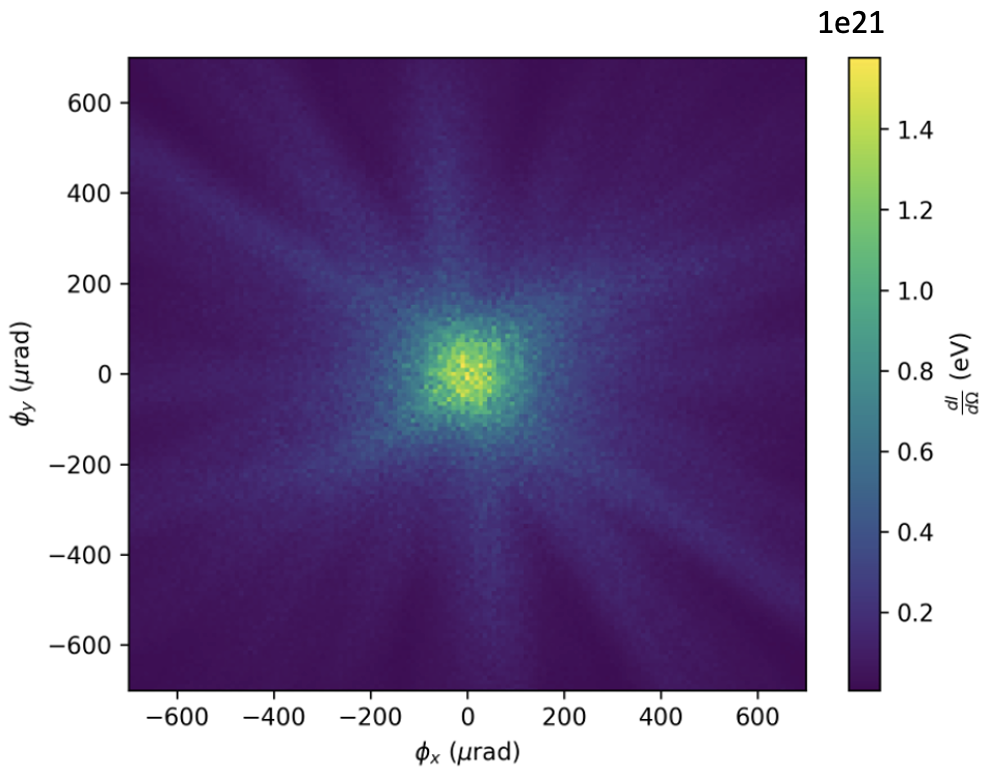}}
   \caption{Betatron radiation angular distributions emitted by the drive bunch.}    \label{fig:angular_drive_radiation_E310}
\end{figure}

The ideal spectrum shown in Fig. \ref{fig:drive_E310_quickpic}(c) is peaked in the range 20-75 keV. The spectrum extends, however, to the 2 MeV range which is approached using gamma spectrometer. The total number of photons is ~E10, and occupy ~mrad in divergence angle, as illustrated in differential spectra shown in Fig. \ref{fig:drive_E310_quickpic}(b). It should be noted that the spectrum will extend to higher photon energy in the case of larger emittance beams, as foreseen in initial runs at FACET-II. 

\begin{figure}
    \centering
\includegraphics[width=\columnwidth]{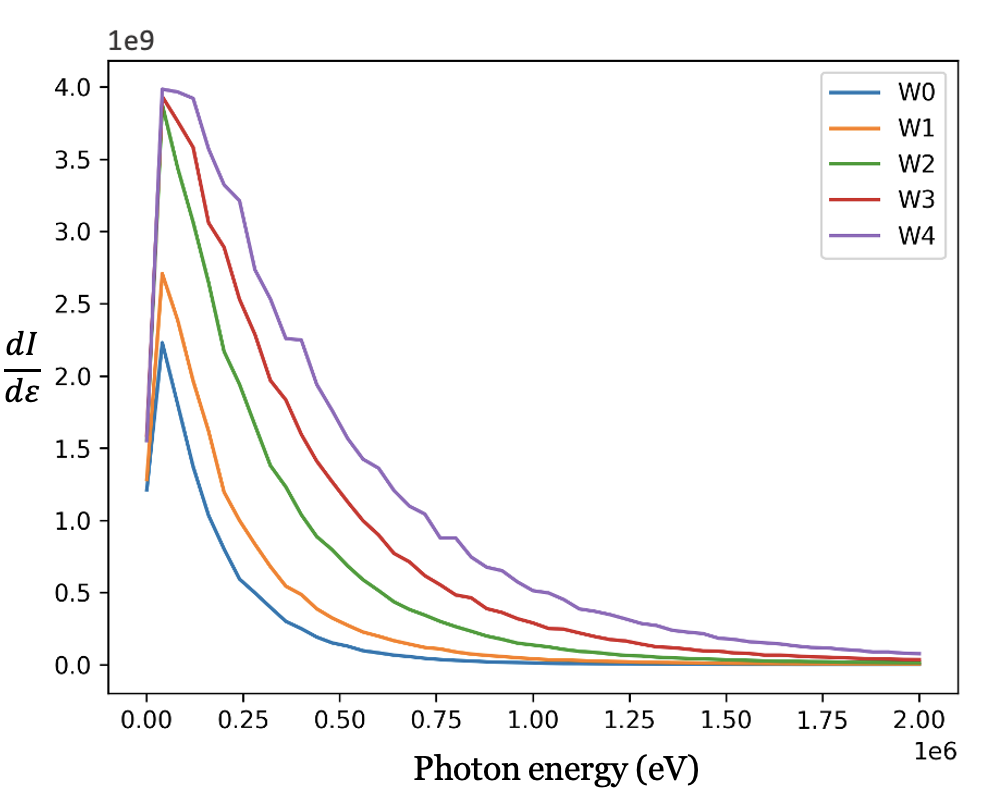}
   \caption{Radiation spectrum of witness beam when  beam is offset for few different cases.} \label{fig:witness_offset}
\end{figure}

In Fig. \ref{fig:angular_drive_radiation_E310}(a), betatron radiation angular distribution and Fig. \ref{fig:angular_drive_radiation_E310}(b) double differential distributions emitted by the drive bunch are shown. There is a spike at the centre as most of the high energy photons are emitted at smaller angles.
Simulations were done for studying different witness offsets effects on the radiation spectrum. In Fig. \ref{fig:witness_offset} a witness beam offset from the axis shows distinct betatron radiation signatures. Several simulations were ran using model \ref{sec:model2}, where the witness beam's $x$ centroid was offset by a value $\Delta x$. The values of $\Delta x$ scanned over were ${0}{\mu m}$, ${5}{\mu m}$, ${10}{\mu m}$, ${15}{\mu m}$, and ${20}{\mu m}$. The $\Delta x = 0$ simulation is the on axis simulation.


\begin{table}[h!]
\caption{Radiation emitted by drive, witness and both beams in PWFA for different witness offsets.}
    \centering
    \begin{tabular}{|c|c|c|}
        \hline 
Simulation & Offsets ($\mu m$) & Witness energy (eV) \\
  \hline   
W0&	0 &4.380e+21 \\
W1&	5 &6.678e+21\\
W2&	10 &1.215e+22 \\
W3&	15 &1.659e+22\\
W4&	20 &2.163e+22\\   \hline      
    \end{tabular}
    \label{tab:offset_wit}
\end{table}


The trailing bunch should be smaller in the longitudinal direction than the drive beam in plasma for loading the wake, and it must contain sufficient charge which can flatten the wake so that the energy spread will be narrow. The drive beam dynamics in these simulations are unchanged. The witness beam dynamics were similar except for two main differences: beams with a larger offset had more emittance growth and $x$ spot size growth. The large offset beams lost some charge, although, for the very most significant offset, it was only $1.21 \%$. Large offset beams had a slightly smaller energy spread with the $\Delta x = {20}{\mu m}$ case having $\Delta \gamma / \gamma \approx 0.711 \%$ compared to the $\Delta x = {0}{\mu m}$ case which had $\Delta \gamma / \gamma \approx 0.856 \%$. In Fig. \ref{fig:witness_offset} spectrum has many variations depending on different offsets of the witness beam. Electrons oscillate at larger amplitudes when witness beam has the highest offset. 

\section{Outlook and future goals} \label{sec:conclusion}

We used different techniques to compute the motion of charged particles, including analytical methods, numerical methods, PIC method, Monte Carlo simulation, and hybrid methods, where we combined two or more of the above techniques to take advantage of their strengths and overcome limitations. We compared and validated our code against the analytic expressions under certain conditions with zero emittance,, finite emittance, and matched spot size. We also minimized witness beam acceleration by placing at wakefield’s zero crossing and the time evolution of the drive beam was turned off. 

Model \ref{sec:model1} is primarily helpful for big data sets. Model \ref{sec:model2} is well-suited for all the FACET-II related experiments and other beam-driven plasma physics because it tells us the exact beam plasma interaction dynamics and radiation emitted. Model \ref{sec:model3} calculates radiation by obtaining trajectories from the full PIC code \texttt{OSIRIS}. We benchmark our codes with \texttt{EPOCH} \ref{EPOCH} which calculates radiation using QED modules that requires information about generated photons. Experimentally analyzing witness beam parameters from betatron radiation is challenging in cases where the drive beam produces most of the radiation. The total radiation emitted by a beam scales as $I_{b,0} \sim Q \gamma^2 \sigma_{\perp}^2$, this is less of an issue for witness beams that have higher charges, larger spot sizes, or higher energies. However our models works well for both higher and lower charge beams.

The full spectrum of betatron radiation produced by the electron beams in the plasma source could be measured using above discussed models. The combined spectral and angular photon spectral yield can provide an indirect measurement of the beam's phase space distribution while inside the plasma, which is critical to fully understanding and ultimately optimizing the beam dynamics inside a PWFA in order to produce high-brightness beams for high energy physics applications. It may also be used to detect deviations from ideal focusing conditions in the plasma, such as what might arise from ion collapse, a potentially severe problem for linear collider beams in a PWFA. Beyond observation and optimization, one may actively seek microscopic control over the beam. As proposed at UCLA, an example is the resonant excitation of betatron oscillations \cite{nathan} via an undulator magnet superimposed on a plasma channel. This amplitude dependence may permit the determination of the beam emittance through a complete frequency spectrum measurement. 


If the oscillations are small ($K_{u}<1$), then one may exploit beam matching condition to measure the emittance. With a Gaussian transverse phase space amplitude distribution assumed, the summation over amplitudes produces an RMS normalized emittance that is remarkably related to the RMS spectral width as $\Delta \lambda_{r m s}=2 \varepsilon_{r m s, n} / \gamma$. If one attempts to measure this spectral width, it is implicit that the radiation emitted should be collected at constant energy $(\gamma)$. This requires a significant length of beam-plasma interaction where ion focusing is present while the acceleration is not. This situation can be obtained using an appropriate plasma profile.

Relevant energies for FACET-II PWFA experiments extend roughly from 1 to 10 GeV, while the plasma density is chosen most often near 10$^{18}$cm$^{-3}$. The relevant photon energies that may be encountered thus range from a few keV to the MeV level, and this must be accommodated through a spectrometer and detector design. It is expected that these $\gamma-$rays will originate from high-energy beams, like the drive beam, that experience a high $K_u$ inside the plasma. For high $K_u$, the spectrum of an individual electron follows a synchrotron-like distribution. Therefore, one must tie the amplitudes to the distribution of critical energies to deduce the emittance through the spectrum. The angular dependence of the spectrum is related to the beam dynamics inside the plasma. For example, an on-axis symmetric bunch performing mismatched oscillations in the plasma will lead to betatron radiation with much higher critical energy in the center than on the side of the photon beam. However, if the beam is matched, there is no correlation between the angle and spectrum in the very high $K_u$ limit. The double differential spectrum can provide simultaneous information regarding matching and betatron amplitudes, which then, in turn, allow constraints on the transverse phase space, particularly the beam emittance. In general, because bunch asymmetries and asynchronous $x$ and $y$ oscillations impact the double differential spectrum, it is essential to rely on LW simulations to interpret and deduce beam parameters and phase space.


Betatron radiation will be an invaluable tool for the diagnostics of upcoming experiments at FACET-II with mobile ions. In these experiments, a long bright beam causes the ions in the blowout bubble to collapse toward the axis producing an ion column \cite{claire}. The strong focusing fields resulting from this will generate large amounts of high-energy betatron radiation. This will lead to a prominent and unique radiation signature that can be used to diagnose the beam-plasma interaction. A flat beam ion motion experiment can demonstrate the formation of an asymmetric beam-ion equilibrium at FACET and AWA. The beam becomes a complex, non-Gaussian distribution due to phase space mixing due to the nonlinear fields, which are a consequence of ion motion. The collisionless relaxation to equilibrium can be seen in the beam spot size and emittance evolution.
The scientific goal of the project is to model codes that can calculate radiation given the information of the particle trajectories for any available beam facilities and then use that radiation spectrum to reconstruct the beam parameters using machine learning algorithms. A complete set of theoretical, computational, and experimental knowledge required for calculating the betatron radiation spectrum is presented.

Recent calculations\cite{benedetti2018giant} have predicted that a high-density ($n_b > 3 \times 10^{19}$ cm$^{-3}$) ultra-relativistic electron beam passing through a millimeter-thick conductor produces bright collimated gamma-ray pulses at high electron-to-photon energy conversion efficiencies, up to 60\%.  This emission occurs by synchrotron radiation in the presence of beam filamentation rather than
ordinary bremsstrahlung. In this collective phenomenon,
characterization of both the angular and spectral properties of the gamma-ray burst at the onset of the beam-filamentation instability is of critical importance.  At FACET-II, the {E-305} experiment ``Beam filamentation and bright gamma-ray bursts'' will study this
phenomenon.

The needs for the FACET-II E-300 collaboration for measuring beam matching are to be able to interpret the integrated signal to reach
the conditions suitable for emittance preservation, the critical goal
of this experiment. There is a strong correlation between minimizing the emittance
growth and observed minimization of the integrated betatron radiation signal. This relationship arises due to the additional radiation emitted when the beam is mismatched.


The radiation diagnostics could also be used in future PWFA-based scenarios for linear collider, where beams with highly asymmetric emittance are expected. Therefore, betatron radiation diagnostic systems are one among those needed for characterizing the beam and will be available at FACET-II. These include the betatron radiation spectrum via a Compton or pair spectrometer, as described in \cite{naranjocompton,Naranjo_1}; the downstream beam imaging systems to determine phase space dilution of accelerated beams in this case, and momentum resolving spectrometers. Betatron radiation models promise functional diagnostics for understanding the beam and plasma interaction dynamics. In addition, betatron radiation will be helpful for the dragon tail experiment at FACET-II \cite{dragontail}. The generated witness beam and the driver beam will produce a betatron radiation signal which can be characterized using above mention betatron models. Once established, the beam-plasma interaction can be interrogated by measuring the betatron radiation spectrum for resonantly excited plasma wakefields in the quasi-nonlinear(QNL) regime. In all of the experiments present at FACET-II, the expected photon spectra will be quite broad, with the exception of the small emittance injected beam
in Trojan horse-like scenarios. We developed robust methods of taking the observed distributions, of the types described in the previous section, to invert the spectral information discussed in \cite{ML-MLE-2022}. This is done using machine-learning algorithms and MLE. This approach requires that the algorithms pick out established patterns in the data.



\section{Acknowledgement}

This work was performed with the support of the US Department of Energy, Division of High Energy Physics, under Contract No. DE-SC0009914, and the STFC Liverpool Centre for Doctoral Training on Data Intensive Science (LIV.DAT) under grant agreement ST/P006752/1. This work used computational and storage services associated with
the SCARF cluster, provided by the STFC Scientific Computing Department, United Kingdom.

\bibliography{references}

\end{document}